\documentclass[11pt]{article}
\usepackage[utf8]{inputenc}
\usepackage{graphicx}
\usepackage{float}
\usepackage{csvsimple}
\usepackage{booktabs}
\usepackage{subcaption}
\usepackage{longtable}
\usepackage[style=authoryear, uniquename=false]{biblatex}
\usepackage{caption}
\usepackage{afterpage}
\usepackage{authblk}
\usepackage{mwe}
\usepackage{adjustbox}
\usepackage{multirow}
\usepackage[bottom]{footmisc}
\usepackage{geometry}
\usepackage{subfiles}
\usepackage{xurl}
\usepackage{amsthm}
\newtheorem{assumption}{Assumption}
\usepackage{amsmath,amssymb}
\DeclareRobustCommand{\bbone}{\text{\usefont{U}{bbold}{m}{n}1}} 
\DeclareMathOperator{\EX}{\mathbb{E}}
\usepackage{setspace}
\title{\vspace{-2cm} Breaking Down the Lockdown: \\ The Causal Effects of Stay-At-Home Mandates on Uncertainty and Sentiments During the COVID-19 Pandemic\footnotetext{We benefited from helpful comments and suggestions from Danielle Braun, Francesca Dominici, Kosuke Imai, Kevin Josey, Fabrizia Mealli, and Davide Viviano and participants at the Harvard Data Science seminar, at the 2022 Institute of Mathematical Statistics (IMS) International Conference on Statistics and Data Science (ICSDS), the 2022 Joint Political Economy and Applied Microeconomics Workshop at the University of Bolzano and at the Advanced Language Processing Winter School (ALPS 2023).}}
\author[1]{Carolina Biliotti\thanks{Co-first authors.}}
\author[2]{Falco J. Bargagli-Stoffi*}
\author[3]{Nicol\`{o} Fraccaroli\thanks{Corresponding author. E-mail: nicolo\_fraccaroli@brown.edu.}}
\author[1]{Michelangelo Puliga}
\author[1]{Massimo Riccaboni}
\affil[1]{\small IMT School for Advanced Studies, Lucca, Italy}
\affil[2]{\small Harvard University, Cambridge MA, USA}
\affil[3]{\small Brown University, Providence RI, USA}

\date{May 2023}
\begin{document} 

\singlespacing
\maketitle

\begin{abstract}
\noindent \small 
We study the causal effects of lockdown measures on uncertainty and sentiment on Twitter. To this end, we exploit the quasi-experimental framework created by the first COVID-19 lockdown in a high-income economy---the unexpected Italian lockdown in February 2020. We measure changes in public sentiment using deep learning and dictionary-based methods on the text of daily tweets geolocated within and near the locked-down areas, before and after the treatment. We classify tweets into four categories---economics, health, politics, and lockdown policy---to examine how the policy affected emotions heterogeneously. Using a staggered difference-in-differences approach, we show that the lockdown did not have a significantly robust impact on economic uncertainty and sentiment. However, the policy came at the price of higher uncertainty on health and politics and more negative political sentiments. These results, which are robust to a battery of robustness tests, show that lockdowns have relevant non-health related implications.
\end{abstract}
\newpage

\doublespacing
\section{Introduction}\label{introduction}

The spread of the SARS-CoV-2 virus and the associated COVID-19 disease was accompanied by an extraordinary rise in uncertainty and negative emotions, resulting in significant economic and social costs. Recent research has shown that during the pandemic, uncertainty peaked and negative emotions surged, explaining much of the decline in industrial production and consumer confidence, as well as the volatility in investment and stock market returns.\footnote{\textcite{Altigetal2020} documented that during the pandemic, indicators of insecurity in the United States and the United Kingdom recorded the highest levels since records began. Estimates show that COVID-19-induced uncertainty resulted in a cumulative 14 percent decline in global industrial production \autocite{Caggiano2020} and explained about half of the COVID-induced decline in U.S. output \autocite{Baker2020}. During the pandemic, consumer confidence declined \autocite{ABOSEDRA2021e00227} and economic sentiment fluctuated \autocite{doi:10.1080/00036846.2022.2061903}, affecting investment volatility and stock market returns \autocite{Çevik, DASH2022101712}. In China, the government's tight restrictions on movement and activity exacerbated negative sentiment among economic actors \autocite{dubey} and fueled political protests that shook financial markets (See, \textit{The Wall Street Journal} (Nov. 28, 2022): \url{https://www.wsj.com/articles/global-stocks-markets-dow-update-11-28-2022-11669635601}). Consistently, the countries that were most affected by the economic downturn were also those that experienced the greatest change in sentiment \autocite {VANDERWIELEN2021105970}. Uncertainty about the duration and extension of the lockdown restrictions and the lack of government commitment have been shown to play a significant role in the negative trend of economic activity during the pandemic \autocite{RePEc:sgh:kaewps:2021060, RePEc:fip:fedmsr:92934}. These findings are consistent with a broad literature highlighting the multiple economic costs of higher uncertainty and worse economic sentiment \autocite{Baker2016, pastor2013political, bruno2015cross, Shapiro2022, Choi2023, Adamsetal2023}. Moreover, the emotional trauma caused by the COVID-19 pandemic had negative social effects, such as psychological distress \autocite{Ferrante2022} and racial hatred \autocite{nguyen2020exploring}.} Against this backdrop, most governments implemented lockdown measures to contain the spread of the virus among their populations.\footnote{Based on a dataset of government responses to COVID-19 for 180 countries, \cite{Haleetal2021} show that by April 2020, lockdown measures had become a widespread policy response to contain the spread of the virus.} Lockdown policies, or stay-at-home mandates, are temporary restrictions that prevent residents from leaving their homes, except for indispensable tasks or for work in essential businesses. While these policies are considered effective in slowing down the spread of COVID-19, as backed by recent evidence\footnote{\textcite{alfano2020efficacy,sen2020association,cauchemez2020lockdown,fowler2021stay, zhang2022evaluating, bodas2022lockdown}.}, their implications on public sentiments remain unclear.

This paper aims to fill this gap and investigate whether lockdowns fuel or mitigate uncertainty and negative sentiment. This question is crucial to better understand whether the economic and social costs of lockdown measures outweigh their health benefits. On the one hand, lockdown measures may have exacerbated already high levels of public uncertainty and negative emotions. Interdiction measures may aggravate uncertainty and negative sentiment by exacerbating health concerns, worsening expectations for the resumption of economic and social activities, and triggering political backlash against the authorities responsible for the restrictions. If this was the case, these policies would add to the already high level of uncertainty and negative sentiment due to the spread of the virus. On the other hand, lockdown measures could prevent the deterioration of public sentiment by signaling commitment and reducing information asymmetry. First, stay-at-home mandates signal the commitment of authorities to contain the spread of the virus, which may reassure the public that the spread of the virus can be contained in the near future. In this case, lockdown measures reduce concern about health conditions, improve expectations for the economy, and generate public support for the authorities. Second, introducing a stay-at-home mandate reduces the information asymmetry associated with suspending economic and social activities. Since, from a theoretical perspective, the effects of lockdowns on public sentiment could go in opposite directions, the question needs to be tested empirically.

However, problems of simultaneity and endogeneity make it difficult to determine the causal effect of lockdowns on public emotions. Because authorities in most countries introduced lockdown measures immediately after the virus was detected, the COVID-19-induced deterioration of public sentiment and the implementation of lockdown measures are concomitant. Endogeneity is also an issue, as lockdown measures were put in place when COVID-19 cases were discovered. These areas are likely the same areas where people perceive a higher risk of pandemic-related costs and where sentiment is therefore lowest. Because of these limitations, most studies on the costs of lockdowns provide evidence that is correlational, not causal.\footnote{See for instance: \textcite{Altigetal2020, Goolsbee2021, Rojas2020}.}

We address these questions by focusing on the case of the February 2020 Italian lockdown, which was the first COVID-19 lockdown in a high-income economy. This case study provides an ideal quasi-experimental setting that allows us to estimate the causal effect of the lockdown policy on public sentiment. The main reason for this is that the lockdown was initially imposed on a few communities where the first individuals to test positive for the virus happened to be detected.\footnote{We provide more details on the quasi-randomness in the assignment of the lockdown policy in Section \ref{Background}.} Nonetheless, the potential transmission of the virus inside and outside the area under lockdown was homogeneous at the time the policy was implemented as shown in epidemiological analyses by \textcite{CEREDA2021100528}. This indicates that the detection of the first cases and the geographic allocation of treatment were random, whereas the incidence of the disease was balanced between treated and untreated units. Health and socioeconomic conditions inside and outside the restricted area were also balanced, suggesting that there was no other inherent difference that increased the likelihood of the first detection of COVID-19 cases in an area over the others.\footnote{See Section \ref{app:random}.} Because this was the first COVID-19 lockdown after China, the introduction of this policy was unexpected, especially in the group of municipalities where it was first introduced. Since at the time, vaccines were not available and testing equipment was lacking, we can better isolate the impact of the lockdown as the only intervention to control the virus during the period studied. The staggered expansion of the lockdown over two main phases allows us to further refine the identification of the causal effect of the policy. While the government imposed the first lockdown on February 23, 2020, for only 11 municipalities, it expanded the mandate on March 8 to a larger area that included 26 provinces. The selection of these provinces was also dictated by the random identification of COVID-19 cases. Finally, on March 9, the government extended the mandate to the entire national territory.

We measure uncertainty and sentiment by combining deep learning and dictionary approaches of natural language processing on the text of Twitter data. By geolocating each account, we can distinguish tweets from users inside the area under lockdown, known as\textit{ red zone}, from those outside. To this end, we collect Twitter data from the red zone and its neighboring cities\footnote{This is the area referred to as the \textit{orange zone} (more on this in the next section). In a placebo test, we also compare tweets in the red zone with tweets across Italy} in the period before and after the imposition of the lockdown. Our sample includes 33,317 unique tweets (2,005 in the red zone; 31,312 outside) from 1,277 accounts. To classify the tweets based on uncertainty and sentiment, we resort to the AlBERTo model, a deep learning natural language processing model based on Bidirectional Encoder Representations from Transformers (BERT) and adapted for textual data in Italian \autocite{polignano2019alberto}.

To better understand the impact of the lockdown on the public's emotions, we divide uncertainty and sentiments into four main dimensions: (i) \textit{economy}, (ii) \textit{health}, (iii) \textit{politics}, and (iv) \textit{lockdown policy}. We consider the first three dimensions based on the growing literature examining how emotions changed after COVID-19. Economic considerations are included since scholars have found that uncertainty and negative expectations about the economy increased substantially during the pandemic \autocite {Baker2020, Caggiano2020, ABOSEDRA2021e00227}. Health considerations are of interest because of the increasing uncertainty and negative sentiment about health brought about by COVID-19 and the heightened health-related risk perceived in the population. The political dimension is relevant as the spread of COVID-19 and the political response of the authorities significantly shaped attitudes toward the incumbent (see \cite{Devine2020} for a review). We add a fourth dimension (which we refer to as \emph{policy} for simplicity) to capture the noise generated by uncertainty and negative feelings associated with the behavioral guidelines of the restrictions. This allows us to control for instances where citizens lacked clarity or dissatisfaction about the impact of policies on daily life, such as school closures, road accessibility, suspension, and restrictions on public activities. We disaggregate the tweets by manually creating a dictionary for each of the four dimensions. We then create a sparse matrix of document features that indicates whether each tweet contains words that appear in each vocabulary.

Using a staggered Difference-in-Differences (DiD) specification, we estimate the causal impact of the lockdown on economic, health, political, and policy uncertainty and negative sentiment. Causal identification is achieved as the \emph{treatment} assignment---the lockdown policy---came as an exogenous shock to the public's uncertainty and negative sentiment, independent of potential outcomes. We adopted a staggered design to account for changes after the initial, geographically limited, lockdown as well as after its expansion to the entire national territory. The results we present are robust to a placebo test and a battery of robustness tests. 

Our findings indicate that the lockdown policy had no significant effect on the public reaction toward economics: indeed, the lockdown was not accompanied by a rise of economic concerns for those subjected to the measures. This finding is relevant in light of the debate over the immediate repercussions of the lockdown on the economy. By contrast, we find that the stay-at-home mandate provoked a change in the health- and politics-related emotions of individuals based in the treated area. Once subject to the lockdown, users express significantly higher uncertainty around health and politics and more negative sentiments around politics than users outside the restricted area (even if, as discussed, exposure to the health risks posed by the virus is comparable in both areas). This indicates that the lockdown generated more concerns by increasing the perception of health-related risk, rather than reassurance by signalling the commitment of the authorities to fight the virus. Notably, this does not necessarily represent a cost. While the spread of uncertainty in the population is certainly not desirable, in the context of COVID-19, it might have been accompanied by higher awareness, and therefore compliance with the measures, which are beneficial to contain the spread of the virus. 

Instead, the increase in negative feelings and, to a lesser extent, uncertainty associated with the policy represents a clear cost that policymakers must bear. Our estimates show that individuals in the treated area express significantly more negative feelings sentiments and uncertainty when they discuss politics after the lockdown than their peers outside the area. These costs could have worrisome implications in the context of future pandemics, as elected officials may be reluctant to implement lockdowns due to the associated political costs, regardless of their effectiveness in containing viruses.

Our results contribute to the ongoing debate on the consequences of lockdown measures, highlighting their unintended consequences and the potential trade-offs associated with such measures. The result related to the economy is useful given the evidence that showed rising uncertainty and negativity during COVID-19 but did not break down how much of this could be attributed to restrictive measures \autocite{Baker2020, VANDERWIELEN2021105970}. We show that while a lockdown could theoretically both inhibit and fuel public concern about the economy, it has no significant overall effects in the Italian case. This does not mean, however, that lockdowns have no impact on the economy: Previous work has shown that stay-at-home mandates have relevant effects on the economy \autocite{Alexander2023}. However, our results show that the economic effects of lockdowns stem from other, more direct channels, such as activity closures, rather than lockdown-induced negative emotions. The results on health concerns are more intuitive, but not trivial. The existing data show that sentiment sentiments on Twitter tended to be more positive during lockdowns, but they do not establish a causal link between the implemented policies and emotions and do not categorize tweets by different topics to see on which topics emotions change \autocite{info:doi/10.2196/19447, Priyadarshini2021}. We show that when the effect of a lockdown is causally identified, the result is reversed. The evidence we present is also noteworthy in light of the burgeoning literature on the effects of restrictions on public attitudes toward politics. Several studies examined the increase in positive attitudes toward incumbent politicians during the pandemic and debated whether this was due solely to a rally effect\footnote{The 'rally effect' or 'rally around the flag' effect describes the increase in positive attitudes toward incumbent parties in the context of a crisis, regardless of the wisdom of their policies. For more details on this effect in the context of COVID-19 see, among others \cite{Devine2020, schraff2021political, Hegewald2022, Kritzinger2021}.} driven by fear, or if part of it was the merit of lockdown measures \autocite{BOL2020, Devine2020, schraff2021political, Hegewald2022, Kritzinger2021, vanderMeer2023}. We show that not only did the lockdown fail to foster positive attitudes toward politics, but that, on the contrary, it actually worsened them. As expected, this represents a clear cost to policymakers, who may risk delaying or weakening their policy response when lockdown measures are necessary. 

Our work also contributes to two separate strands of the literature. First, we contribute to the growing literature on evaluating the impact of lockdown measures. Several studies have shown that lockdown policies have been proven to be effective in limiting the spread of the virus and reducing the burden on hospitals and medical facilities \autocite{cauchemez2020lockdown, Melnickm1924, Froste044149, Cho2020, Cerqueti2021, Sjödin2020}. Others documented their impact on non-health related aspects, such as lower consumption \autocite{Alexander2023}, higher inequality and economic strain \autocite{palomino2020wage}, poorer mental health conditions \autocite{banks2020mental,elmer2020students}, increased doom-scrolling  and addiction to social media \autocite{anand2022doomsurfing}, political support and public trust \autocite{BOL2020, Devine2020, schraff2021political, Hegewald2022, DeVries2021, Kritzinger2021, vanderMeer2023, Bonotti2021, Fazio2021political}. We complement these works by focusing on the causal impact of lockdown measures on public emotions, for which, to our knowledge, there is only correlational evidence \autocite{info:doi/10.2196/19447, Priyadarshini2021, BOL2020, Hegewald2022}. Understanding the causal impact of imposing a stay-at-home mandate on public emotions is critical for policymakers to thoroughly evaluate the benefits and costs of such policies.

We also add to the literature on uncertainty and sentiment, which generally relies on nontraditional data such as textual data. Large unexpected economic, political, and social shocks increase uncertainty, distress, and confusion among the public \autocite{Murray346, Baker2016, 10.1145/3025453.3025891, NBERw26243}. Recent studies have documented particularly high levels of economic uncertainty \autocite{YANG2021101597} and health-related negative emotions \autocite{info:doi/10.2196/19447} following the enactment of COVID-19 lockdown policies, as well as increased political polarization \autocite{https://doi.org/10.1002/hbe2.202, 10.3389/fpos.2021.622512}. Yet there is less clarity about what particular aspects have contributed to bringing about such changes. We contribute to filling this knowledge gap by examining how lockdowns can affect uncertainty and sentiment on various issues. Our work contributes to this literature by showing that uncertainty and negative sentiment did not increase uniformly across subjects after lockdown, but rather followed different dynamics. Thus, this is the first study to analyze the effects of stay-at-home mandates on multiple dimensions of uncertainty and sentiments.

The rest of the article is organized as follows. Section \ref{Background} provides an overview of the timing of the Italian lockdown and argues about the random choice of the area under lockdown. Section \ref{sec:methods} presents the methods we use to generate our data and conduct our empirical analyses. First, in Section \ref{ml}, we describe in detail the machine learning model we implement to compute uncertainty and sentiment based on Twitter text data. Second, in Section \ref{empirical}, we introduce the empirical strategy and the main econometric specification. More detailed information on the methodology is provided in the Online Appendix. Section \ref{data} describes the data collection and classification process. In Section \ref{results} we present and discuss the estimation results and a series of robustness tests. The final section concludes.

\section{Background: the Italian COVID-19 lockdowns}\label{Background}

In this section, we provide a brief overview of the timeline of the Italian lockdown to show that, conditional on the region of Northern Italy under investigation (see Figure \ref{fig:map_zones}), the assignment to treatment---i.e. the allocation to the lockdown of certain areas instead than others---was random and unexpected. The brief background we provide in this section is helpful to motivate the research design we adopt. The empirical strategy is comprehensively described in Section \ref{empirical}.  

\subsection{Timeline of the Italian lockdowns}

Italy was the first Western country, and the second country in the world after China, to detect a rapid spread of the COVID-19 virus on its territory. In response, the government imposed lockdown measures in some specific geographic areas on February 22, 2020. The selection of the area affected by the lockdown, which was named\textit{red zone}, was based on the (accidental) detection of the first COVID-19 cases, which happened to be registered in a specific area in northern Italy. These restrictions were then extended in two phases, the last of which was announced on March 10, 2020, and covered the entire national territory. Italy's lockdown anticipated similar measures introduced in other countries starting in the second half of March 2020 (\cite{Haleetal2021}). The peculiarity of the staggered introduction of the lockdown policy in northern Italy is that the allocation to geographic areas was unintentionally random, as it depended on the random identification of the first positive cases. Given the unexpected and rapid spread of the virus, the authorities were not aware at the time of the introduction of the measure that there were a comparable number of COVID-19 cases in other areas just outside the red zone.
On February 20, 2020, the first-ever case of COVID-19 infection was registered in Italy. After meeting a friend returning from a trip to China, a 38-year-old man from the town of Castiglione d'Adda, a small municipality in northern Italy in the Lombardy region, visited the hospital in the nearby town of Codogno on February 18, 2020, with flu symptoms. The hospital soon discharged him and the man returned to his hometown.\footnote{See,\textit{Il Giorno} (October 28, 2021): %
\url{https://www.ilgiorno.it/lodi/cronaca/paziente-1-covid-1.6969895}.} The following day, the same person was admitted to the intensive care unit of Codogno Hospital with pneumonia and later tested positive for COVID-19. The discovery of the first COVID-19-positive case was incidental, as the decision to test the patient for coronavirus was based on the hunch of one of the physicians.\footnote{See, \textit{Ansa} (Feb. 20, 2021): %
\url{https://www.ansa.it/sito/notizie/speciali/editoriali/2020/02/21/speciale-coronavirus-un-anno-di-pandemia-tutto-comincio-a-codogno_170ab350-a59d-4b34-9095-6dc9cb6def06.html}.}

Within 24 hours, several individuals in contact with the first patients were tested and found positive for COVID-19. This was the first confirmed case of transmission of the new virus and the epicenter of the COVID-19 outbreak in Europe. In the days that followed, more and more people in the area tested positive for COVID-19.\footnote{See, \textit{la Repubblica} (Feb. 21, 2020):\url{https://milano.repubblica.it/cronaca/2020/02/21/news/coronavirus_a_milano_contaggiato_38enne_e_un_italiano_ricoverato_a_codogno-249121707/}.}

\begin{figure}[H]
\centering
\includegraphics[scale = 0.20]{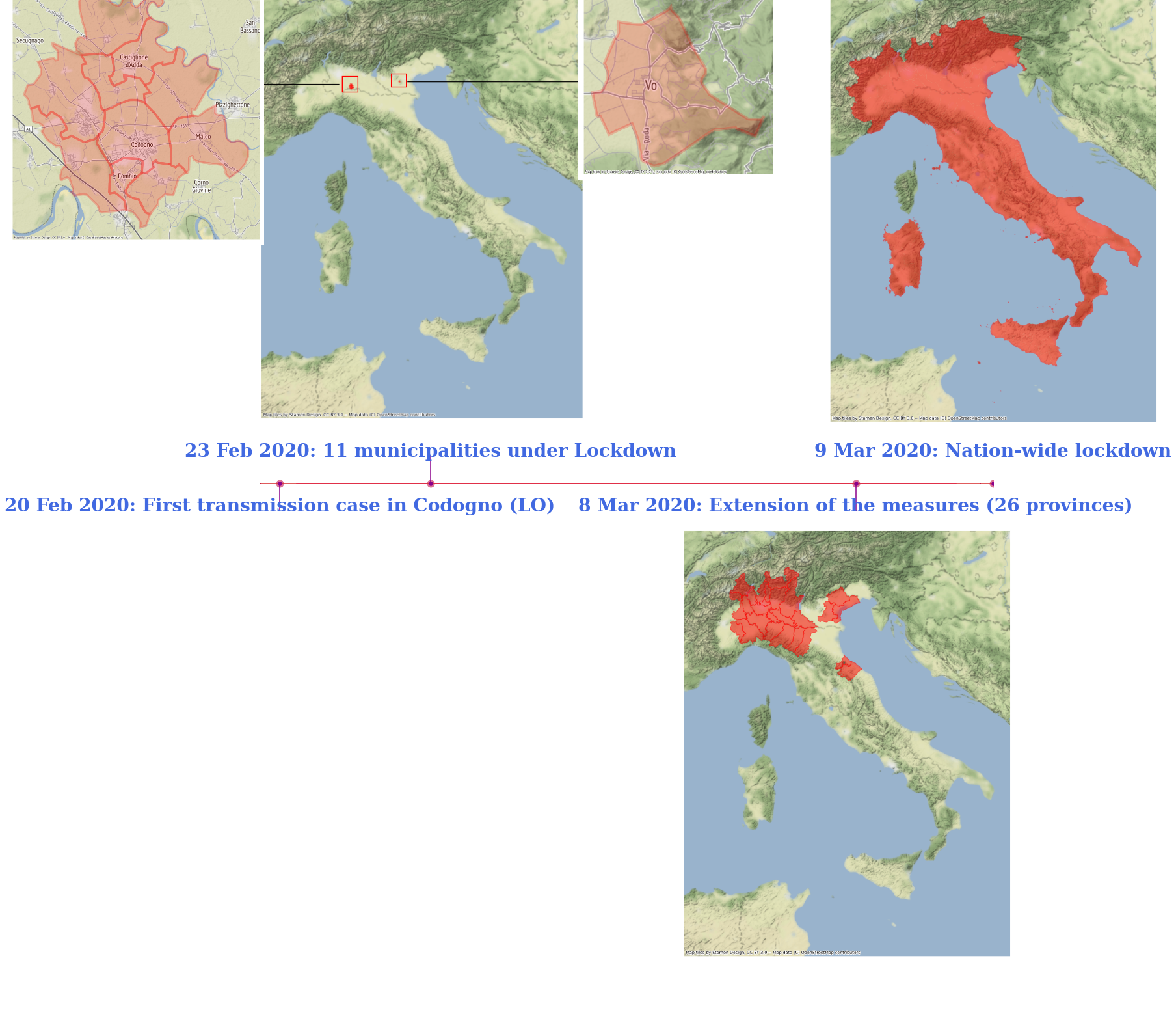}
\caption{Evolution of the lockdown measures in Italy.}
\label{fig:timeline}
\end{figure}

On Feb. 21, in the town of Codogno, as well as in nine neighboring municipalities (Castiglione d'Adda, Casalpusterlengo, Fombio, Maleo, Somaglia, Bertonico, Terranova dei Passerini, Castelgerundo e San Fiorano), all public gatherings, non-essential businesses, and indoor activities were suspended.\footnote{See, \textit{Quotidiano Sanità} (Feb. 2, 2020): %
 \url{http://www.quotidianosanita.it/scienza-e-farmaci/articolo.php?articolo_id81591}.} The lockdown specifically targeted the hometown of the first patient discovered and the cities where his closest contacts were located. On February 22, the Italian government announced a new Decree\footnote{See, \textit{Gazzetta Ufficiale} n.45 (Feb. 23, 2020):
\url{https://www.gazzettaufficiale.it/eli/id/2020/03/09/20A01521/sg}.}
which imposed strict quarantine measures on the ten cities in the Lombardy region starting the next day (February 23, 2020).\footnote{Authorities also imposed a lockdown in Vo' Euganeo (Veneto), where a deceased 77-year-old man tested positive for COVID-19 on February 21. The man had no direct or indirect contact with China, the key element for a COVID-19 diagnosis at the time. The case of the small town of Vo' Euganeo (about 3\,200 inhabitants) represents an isolated case that was not included in our analysis because it is not considered part of the first Italian epicenter identified in the Codogno area} affected by the pandemic. This lockdown led to the forced isolation of more than 50,000 people in the area highlighted in Figure \ref{fig:timeline} (upper left panel). Authorities referred to these cities as the \emph{red zone} (\emph{zona rossa} in Italian). An \emph{orange zone} (\emph{zona arancione}) was established to cover the municipalities near the red zone (see Figure\ref{fig:map_zones} for an example). The orange zone was subject to milder restrictions such as school closures and the suspension of all gatherings (e.g., theaters and religious institutions), while limited bar and restaurant hours and transit were allowed within the zone.

The lockdown in the red zone was enforced by the national police and gendarmerie (Carabinieri), which reportedly deployed 400 units at 35 checkpoints.\footnote{See, \textit{Quotidiano Sanità} (Feb. 2, 2020):\url{http://www.quotidianosanita.it/scienza-e-farmaci/articolo.php?articolo_id81591}.} The mission of these units was to enforce the lockdown measures by blocking the inflow (outflow) of people into (out of) the quarantined area. Red Zone residents were allowed to leave their homes only for essential supplies such as food, medicine, and sanitation. Face masks were mandatory when they were away from home. Authorities prohibited in-person attendance at schools, workplaces, and public gatherings. Trains also bypassed the restricted area.\footnote{See, \textit{Gazzetta Ufficiale} n.45 (Feb. 23, 2020):
\url{https://www.gazzettaufficiale.it/eli/id/2020/03/09/20A01521/sg}}

\vspace{0.5cm}

On March 7, the government announced the extension of the \emph{red zone} to all of Lombardy and 14 other cities outside that region (Modena, Parma, Piacenza, Reggio nell'Emilia, Rimini, Pesaro e Urbino, Alessandria, Asti, Novara, Verbano-Cusio-Ossola, Vercelli, Padua, Treviso, Venezia), to take effect the following day, placing a total of 16 million people under a strict stay-at-home mandate.\footnote{See, \textit{Gazzetta Ufficiale} (Mar. 8, 2020):
\url{https://www.gazzettaufficiale.it/eli/id/2020/03/08/20A01522/sg}.} Orders implementing the restrictions nationwide were announced on March 9.\footnote{See, \textit{Gazzetta Ufficiale} (Mar. 9, 2020):\url{https://www.gazzettaufficiale.it/eli/id/2020/03/09/20G00030/sg}} Around 8 p.m., Italian Prime Minister Giuseppe Conte described the seriousness of the situation in a televised announcement and declared the extension of the lockdown to the entire national territory starting March 10, 2020. The intensity of the stay-at-home measures loosened up only in May and the lockdown was reintroduced during the next major outbreak of the winter season.

\subsection{Random choice of the area under lockdown}\label{sec:random}

We argue that the discovery of the first COVID-19 case in Codogno, and not in another municipality of the studied area encompassing the red and orange zones (see Figure \ref{fig:map_zones}), was simply a result of chance. The identification of the first patient happened by chance in this particular municipality, since other municipalities (not included in the first lockdown ) were exposed to the virus as well as Codogno, as evidenced by epidemiological research \autocite{CEREDA2021100528}. When physicians learned that the patient, who showed symptoms attributable to COVID-19, had contact with a friend who had recently been on a business trip to China, they decided to force the existing protocol, of testing only Italians and foreigners who had been in China.\footnote{See, \textit{La Stampa} (Feb. 20, 2022):
\url{https://www.lastampa.it/cronaca/2022/02/20/news/codogno_20_febbraio_il_tampone_0_che_scopri_il_covid_in_italia_la_dottoressa_annalisa_malara_ricordo_ancora_la_paura_or-2859352/}.}

As a consequence of the first positive case, hospitals began testing people for COVID-19 in the area of Codogno. This decision was made simply by identifying the first COVID-19 patient in Codogno. In fact, recent retrospective epidemiological studies on the transmission risk of coronavirus in Lombardy in February 2020 revealed a homogeneous transmission potential of the virus in the different provinces of Lombardy at the time of the discovery of the first case in Codogno \autocite{CEREDA2021100528}. This means that the disease incidence was balanced between the towns in the red zone and those in the orange zone. To further check the random allocation of the lockdown, we test for differences between the red and orange zones based on their observable characteristics. If there are no differences, this provides support for random allocation. The results of these tests are reported in Section \ref{app:random} of the Appendix. No statistically detectable differences are found. 

Despite these similarities, the lockdown measures were imposed only in the red zone until March. Therefore, we can assume that the treatment was the only difference between the two areas.

\section{Methods}\label{sec:methods}

Our goal is to estimate the impact of the lockdown on uncertainty and negative sentiment. To this end, we first use machine learning to measure the emotions expressed in Twitter messages (or tweets), drawing on Natural Language Processing (NLP) techniques. We then use the emotion scores we compute as outcome variables in the causal model to determine the lockdown's impact on uncertainty and sentiment. Following this structure, in this Section, we first introduce the models used to classify our response variables (Sections \ref{ml} and \ref{dictionary}). We then explain the causal inference methodology used to analyze the data and draw causal conclusions (Section \ref{empirical}).

\subsection{Machine Learning for uncertainty and sentiment measurement}\label{ml}

We measure the uncertainty and sentiment in each tweet using a Bidirectional Encoder Representations from Transformers (BERT). BERT is an advanced open-source transformer model for classifying text data, first developed at Google by \textcite{ https://doi.org/10.48550/arxiv.1810.04805}. One of the key innovations of BERT is the masked learning pre-training objective. In pre-training, the model is exposed to a large amount of unlabeled text data to learn the patterns and structures of the language. In this phase, BERT is trained to predict missing or masked words in a sentence. This is done by randomly masking certain words in the input sentence and then training the model to predict the masked words based on the context.\footnote{The masked learning pre-training objective serves as a form of self-supervised learning, where the model learns from the data provided to it without explicit labels \autocite{https://doi.org/10.48550/arxiv.1810.04805}. By attempting to predict the missing words, the model is forced to capture the relationships between words, grammar rules, and the overall context of the sentence.}

To perform the masked learning pre-training objective BERT uses bidirectional attention. This means that the model learns the contextual, positional, and semantic representation of the term by considering both sides of the masked term. The text is scanned for meaning and context in both directions: first, from left to right, as any other transformer would, but then also from right to left.\footnote{For fine-tuning specific tasks, the model is initialized with pre-train parameters, and for each task, we plug in a classification layer and train end-to-end with labeled inputs.} This feature is a significant advantage over existing strategies that measure uncertainty and sentiment based on the simple matching of a set of words. For example, the popular index developed in \textcite{Baker2016} measures economic uncertainty by calculating the co-occurrence of the word 'uncertainty' or 'uncertain' and a set of words related to the economy (researchers used the same bag-of-words approach to measure sentiment as discussed in \textcite{Shapiro2022}). However, such an approach does not consider the text surrounding the word uncertainty, including the words that precede or follow it. BERT overcomes this gap by scanning the surrounding text in both directions. In addition, by using pre-trained text, we can detect different degrees of uncertainty and negative emotion. Thus, we can examine emotions beyond a unidimensional setting.

BERT efficiently detects and distinguishes when words are used in different contexts. As a transformer-encoder, BERT has multiple self-attention layers, where words are transformed into highly contextualized numerical representations. What distinguishes BERT from other pre-trained transformers used for fine-tuning, such as OpenAI's GPT \autocite{alt2019finetuning}, is its bidirectional attention, as GPT is only allowed to attend to the left context of a word. The better the model understands the different usage and meaning of words, the better it can then perform in classification \autocite{Metzler}. Unlike bidirectional Long Short-Term Memory (LSTM) models such as ELMO \autocite{elmo} for unsupervised learning, the attention mechanisms ensure that the model is able to handle long-term dependencies when processing long sequences such as when dealing with text.

During pre-training, BERT learns to extract high-quality features from a very large corpus of documents of general nature (e.g., an online encyclopedia). Fine-tuning the initialized parameters of the pre-trained BERT in a downstream classification task allows the model to adapt its parameters to the specific domain of our data sample.

Since the text of the tweets we analyzed is in Italian, we tune the pre-trained AlBERTo model \autocite{polignano2019alberto} to downstream binary classification tasks. AlBERTo is a Deep Learning natural language model trained on a large corpus of tweets in Italian ($\sim$200 million). The model mimics the architecture of BERT and provides a multi-layered bidirectional transformer encoder that understands the Italian language. Following BERT, AlBERTo was also pre-trained with a masked learning task.

When confronted with benchmarks, AlBERTo achieves top results on several NLP tasks, such as sentiment analysis (subjectivity, polarity, and irony classification) and hate speech detection \autocite{polignano2019alberto}. In addition, BERT-based models have proven to be a powerful tool in analyzing public opinions and sentiments expressed on Twitter during the COVID-19 pandemic\autocite{BLANCO2022102918, 10.31234/osf.io/xk5hz, idr13020032}. This includes detecting fake news or bots \autocite{9803414, 9666618} and even anti-vaccine tweets \autocite{ijerph18084069}.

Formally, we can express the masked learning model objective function as:\footnote{For a discussion, see \textcite{Mishev2020EvaluationOS}.}
\begin{equation}
\max\limits_{\theta} \; log(p_{\theta}(\bar{x}|\hat{x})) \approx \sum_{s}^{S}m_{s} log(p_{\theta}(x_{s}|\hat{x})) = \sum_{s}^{S}m_{s} log \frac{ exp(H_{\theta}(\hat{x})_{s} e(x_{s})) }{ \sum_{x'} exp(H_{\theta}(\hat{x})_{s} e(x')))},
\end{equation}
where $x = [x_{1}, ..., x_{S}]$ is the text sequence, $\bar{x}$ the sequence of masked tokens\footnote{The masking rate is the set to be 15\% in accordance with the literature \autocite{https://doi.org/10.48550/arxiv.1810.04805}.},$\hat{x}$ the text sequence with masked tokens and $p_{\theta}$ is the approximation of $p_{\theta}(\bar{x}, \hat{x})$, which is made possible as masked tokens are mutually independent. $m_{s}$ is an indicator function for whether token $s$ is masked and $H_{\theta}$ is the transformer model. 

Fine-tuning is done by adding an additional dense layer after the last hidden state of the classification token $C \in \mathbb{R}^{Z}$, where $Z$ represents the size of the hidden states. The classification objective function is given by
\begin{equation}\label{softmax}
    o = log(softmax(CW^{T})).
\end{equation}
The weights $W \in \mathbb{R} ^{K\times Z}$, with K being the number of labels, are the only new parameters introduced during fine-tuning.

Our approach is to refine the pre-trained AlBERTo model to classify tweets according to reactions detected in the text, such as uncertainty and negative sentiment. We set the training objective to the equation (\ref{softmax}) and train the model on a subset of observations ($\sim 20\%$) provided with \emph{manually} assigned labels, indicating possible degrees of uncertainty (\emph{medium}, \emph{low} or \emph{high} uncertainty) and sentiments (\emph{medium}, \emph{positive} or \emph{negative})\footnote{Authors 1, 2, and 3 manually assigned a subset of tweets to labels. The labels were then independently validated and verified by the three authors}. 

We follow a procedure consisting of two steps to learn the classes of unlabeled text and identify the tweets that express uncertainty or negative sentiment ---i.e., the response of interest in our analysis. For either uncertainty or sentiment, in the first step, we tune the model on all manually labeled examples to perform a binary downstream task where the model learns how to efficiently distinguish the \emph{medium} class--- the neutral tweets--- from the rest. This fine-tuned model is used to classify the reaction in tweets without labels as either neutral (\emph{medium}) or \emph{not} neutral. In the second step, we then use only manually tagged tweets expressing non-neutral reactions as input to an additional binary classification task, where we fine-tune AlBERTo to classify sentences as \emph{uncertainty} or \emph{certainty} (high vs low uncertainty), or distinguish between \emph{positive} and \emph{negative} sentiments. This second fine-tuned model is used to predict the final label of the non-manually assigned tweets that were detected as non-neutral in the previous step. In summary, each step of the process is treated as a separate binary downstream task for fine-tuning, as we re-train the classifier at each step and for both dimensions of public reaction of interest--- i.e., uncertainty and sentiment. Since the procedure involves two steps and the emotions of interest are uncertainty and sentiment, we fine-tune the model using four different binary classification tasks.

We follow this approach instead of proceeding directly with multi-label classification because fine-tuning the Alberto model directly in a multi-label classification task performs much worse than the two-step procedure. The fine-tuned AlBERTo model can more efficiently detect and distinguish classes that are highly separated from each other, e.g., in identifying neutral tweets from the rest or in distinguishing between \emph{positive} and \emph{negative} sentiments. We also perform a hyperparameter grid search for each classification step of the procedure to select the best model for each classification task.\footnote{For a detailed description of the AlBERTo model, the training process, and the out-of-sample performance of the model on a test subsample, see the Appendix Section \ref{cls_model}}

\subsubsection{Classification of uncertainty and negative sentiment with ALBERTo}\label{data_unc}

We measure uncertainty and negative sentiments using the pre-trained AlBERTo on the TWITA dataset.\footnote{The dataset is available on GitHub. See, \url{https://github.com/marcopoli/AlBERTo-it}.} We fine-tune the model to perform separate binary classification tasks, to distinguish and identify the tweets expressing uncertainty and negative sentiments, rather than indifference, positivity, and certainty. 
 
We first pre-process all tweets, as particular characters like hashtags, emoticons, URLs, phone numbers, and dates are tagged with fixed tuples. We remove punctuation and non-UTF symbols. Then, we tokenize tweets, dividing the text into unique words or sub-words according to a large vocabulary of 128,000 terms from the TWITA dataset.\footnote{We use the \texttt{SentencePiece} algorithm \autocite{polignano2019alberto}.} We set 128 as the maximum sequence length for each input sequence of tokens. Naturally, the vocabulary will ignore non-existing words in 2019, such as the term \emph{COVID}, which will be inevitably converted into two separate tokens, \emph{cov} and \emph{id}. This may represent an obstacle for left-to-right transformer models, where each token can only attend to the previous one. On the other hand, BERT provides a bidirectional representation of terms, as the context, semantics, and position of each term are derived by scanning the text in both directions. Therefore, the model will be able to learn the correct context every time a subword like \emph{cov} comes up, considering every other word in the text, and their positioning, to derive contextualization. 

After pre-processing, we process tokenize the input text---i.e., we turn it into a sequence of \emph{tokens}. The text is segmented into words and sub-words, the tokens, that are matched to a large vocabulary. If a sub-word is not recovered in the vocabulary, it will be split into single characters. All sub-words are annotated with a $\#\#$. The first token of an input example is always a classification token, \texttt{[CLS]}, used as aggregated sequence representation for classification tasks. If examples are in pairs---such as \texttt{(Question,Answer)}---a special separation token will distinguish the two sentences \texttt{[SEP]}. To get numerical representations of tokens, each token is converted to its \emph{learned} vector embedding of size 768, a vector representation for each word in the vocabulary that is usually retrieved during pre-training. The final input representation will consist of the sum of the token embedding, position embedding, and sentence-position embedding.\footnote{More details are provided in Section \ref{cls_model} of the Appendix.} 

After fine-tuning the model and predicting the labels of all tweets via deep learning, we obtain two binary variables named \emph{uncertainty} and \emph{negative sentiment} for each tweet in our dataset. The variable \emph{uncertainty} will equal 1 if the tweet expresses uncertainty, zero if the tweets \emph{does not} express uncertainty---i.e., the text is conveying a different emotional state, like certainty or indifference. The same reasoning applies to the \emph{negative sentiment} categorical variable: it is set to 1 if the tweet expresses negative sentiments, zero if the text \emph{does not} express negative sentiment---i.e., it is expressing positivity or indifference.

 \subsection{Dictionary-based classification for topics}\label{dictionary}

Our aim is to identify how uncertainty and negative sentiments changed across different topics once the lockdown is implemented. For this reason, we select four categories that are likely to be influenced by the policy: \emph{health}, \emph{economics}, \emph{politics}, and \emph{lockdown policy} (henceforth referred to as \emph{policy} for simplicity).

These categories are relevant for different reasons. The inclusion of the health dimension of uncertainty and sentiments is straightforward since lockdown measures aim to reduce the population's concerns about the emergency and safeguard the health of the population. We then include the economic dimension, since economic costs are the most direct drawback related to lockdowns due to the suspension of major productive activities. This motivates the inclusion of the economic topic, which will identify when uncertainty and negative sentiments are related to concerns around the economy. Identifying health and economic-related tweets is crucial in order to understand how lockdown policies differently affected the perception of health and economic risks and investigate the health-economic trade-off \autocite{doi:10.1080/01402382.2021.1933310}. 

There is a third dimension of potential distress that is political: since the authorities ultimately responsible for the lockdown are politicians, this could give rise to negative feelings that are of political nature. While political uncertainty can have economic implications, the literature has analyzed this dimension of uncertainty (and sentiments) separately from economic uncertainty due to its different sources and implications \autocite{pastor2013political}. Lockdowns can also affect sentiments toward politicians, but the evidence is mixed. While some findings indicate that lockdowns increase trust in the government \autocite{BOL2020, OudeGroeniger2021}, others found that lockdowns have no effect on trust in the government and that the fear of the pandemic, not the lockdown, is responsible for the improvement in political trust \autocite{Hegewald2022, schraff2021political, Baekgaard2020}. Interestingly, \textcite{DeVries2021} show that when the Italian lockdown was implemented, trust in the incumbent improved also in the neighboring countries that had not yet implemented the lockdown. This suggests that it is important to compare the effect of the lockdown among treated and non-treated individuals. 

Finally, we include a fourth category capturing uncertainty and sentiments associated with the policy itself. This category allows us to distinguish whether rising uncertainty and negative sentiments are associated with the authorities implementing them (political uncertainty) or with the policy itself (uncertainty around lockdown policies). For instance, uncertainty may grow because affected citizens find the rules surrounding the policy unclear.

We categorize tweets into each of these four groups to pinpoint the specific areas where uncertainty and negative sentiments tended to cluster in response to the policy. By classifying each tweet to one of these topics, we can investigate, for instance, whether lockdown measures fuelled higher uncertainty around the economy than around the health situation. To this end, we rely on a dictionary-based classifier that defines for each topic a list of topic-specific commonly used words. The classifier then counts the number of terms used in each tweet that belongs to each list.\footnote{The complete list of words used to construct the topic-specific dictionaries is reported in the Appendix, Section \ref{topics_appendix}.} 

Finally, we create a dummy variable, one for each dictionary---i.e., topic, that equals 1 when the tweet contains at least one term from the dictionaries and 0 otherwise. We allow topic labels to overlap in the same tweet, as tweets may cover multiple topics at once. For example, the binary topic variable for economics is set to 1 if the tweet contains at least one word from the economics dictionary and 0 if the tweet features no words from the dictionary. 

In the process of extracting topics via dictionary-based classifiers, we clean the tweets and remove all URLs, tagging, and punctuation. For hashtags, we remove the \# but keep the tagged word. We drop a total of 153 tweets that featured only tags and ats and ended up being empty after processing. In Section \ref{sec:top_tweets} of the Appendix, we show a sample of tweets with the highest Shannon entropy by emotion-topic pair and discuss the performance of the machine learning classifier in identifying tweets related to each topic and emotion. 

\subsection{Causal modeling}\label{empirical}

Our goal is to measure the effect of the lockdown on uncertainty and negative sentiments. To this end, we exploit the exogenous shock of the lockdown on the treated area. The lockdown was unanticipated and individuals were randomly allocated to the lockdown (aka the treatment). To estimate the effect of the lockdown we adopt a staggered DiD approach, which we describe in this section.

Let $i$ be a tweet from user $j$. Let $T$ be the time period of analysis with $T_{ij} = t$ for $t$ periods, with $t \in \{0,1,2\}$---i.e., before ($t=0$) and after ($t=1$) the implementation of the lockdown of February 23, 2020, and following the nationwide measures of March 9, 2020 ($t=2$). Users are assigned to the targeted lockdown in period $t=1$, while at $t=2$ all users become treated. We define $D_{ij}$ as the \emph{treatment status} indicator: $D_{ij}=1$ if tweet $i$ is from a user $j$ who is in the area subject to the lockdown of February 23---i.e. the \emph{treatment}---$D_{ij}=0$ otherwise. 

Formally, we estimate the following DiD equation on repeated cross-sections of tweets:
\begin{equation}\label{did_model}
Y_{ij,t} = \alpha + \beta_j  + \sum_{t \in \{1,2\}}\lambda_{t}\bbone[T_{ij} = t] + \sum_{t \in \{1,2\}}\delta_{t}(\bbone[T_{ij} = t] \times D_{ij}) + \epsilon_{ij,t} 
\end{equation}
where $Y_{ij,t}$ is the binary outcome variable obtained from the NLP classification presented in the previous section (i.e., uncertainty and negative sentiments, aggregated and grouped by topics) observed for unit $i$ observed in period $t \in \{0,1,2\}$. The time period dummies $\lambda_{t}$ indicate the pre-post-post lockdown periods. Our coefficient of interest is $\delta_{1}$. $\beta_j$ is a user-level fixed effect.

In this DiD setting, the average treatment effect on the treated in period 1, denoted as $\text{ATT}(1)$, is the causal estimand of interest. It is defined as
\begin{equation}
    \text{ATT}(1) = \EX \left[Y_{ij,1}(1) - Y_{ij,1}(0) \,|\, D_{ij} = 1\right],
\end{equation}
where $Y_{ij,t}(0)$ represents the potential outcome of tweet $i$ for user $j$ in period $t$ if they remain untreated at time $t$, and $Y_{ij,t}(1)$ represents the potential outcome at time $t$ if they were treated at time $t$. For each unit, we can only observe one of the potential outcomes at each time period.

Under the assumptions of (i) parallel trends, (ii) no anticipation of the treatment, and (iii) stable unit treatment values assumption \autocite{roth2023whats, RePEc:arx:papers:2105.03737}, the average treatment effect on the treated in period 1 can be rewritten as:
\begin{equation}
    ATT(1)  = \EX[Y_{ij,1} - Y_{ij,0} | D_{ij} = 1] - \EX[Y_{ij,1} - Y_{ij,0} | D_{ij} = 0]
\end{equation}
provided that units are drawn at random from a larger population. These assumptions are discussed in more detail in Section \ref{sec:assumptions_did} of the Appendix.

As the control units eventually become treated once the measures are extended nationwide, our setting features a \emph{staggered} treatment timing, where units receive treatment at different time periods. Following \cite{CALLAWAY2021200}, we let $G$ be a time period indicator when a unit starts treatment. Define $G_{ij,g}=\bbone[G_{i,j}=g]$, a binary variable indicating whether tweet $i$ comes from a user $j$ who received the treatment for the first time in period $g$, and $D_{ij,t}=\bbone[T = t] \times D_{ij}$, indicating whether tweet $i$ is from user $j$ who is treated at time $t$. 

When the identifying assumptions of no treatment anticipation and parallel trends based on the \emph{not-yet treated groups} hold unconditionally on covariates, the causal estimand of the average treatment effect is given by:
\begin{equation}\label{ATT(g,t)}
    ATT(g,t) =  \EX[Y_{ij,t} - Y_{ij,g-1} | G_{ij,g} = 1] - \EX[Y_{ij,t} - Y_{ij,g-1} | D_{ij,t} = 0]
\end{equation}
for $g \leq t < \hat{g} = max_{1,...,n}\{G_{ij}=g\}$.---i.e. for $t$ between the last period in which a new group enters treatment and the time of first entering treatment\footnote{We refer to \cite{CALLAWAY2021200} for details on the generalization of assumptions of the DiD model in a staggered setting.}. This estimator measures the changes in outcomes between the current period and the first period preceding the treatment ($g-1$), comparing units firstly treated at time $g$ and those yet-to-be treated at time $t$. This means that $ATT(g,t)$ cannot be identified for time periods after the very last treatment group starts the treatment, nor we can identify the treatment effect for this last treatment group, as we lose any counterfactual quantities. 

The estimand in \eqref{ATT(g,t)} generalizes the standard two-group, two-periods DiD estimator, where the not-yet-treated group is taken as a suitable counterfactual group. In our setting, $t \in \{0,1,2\}$, $\hat{g} = 2$ and the time reference is $t = 0$. Then it follows that
\begin{equation}
    ATT(1,1) =  \EX[Y_{ij,1} - Y_{ij,0} | G_{ij,g} = 1] - \EX[Y_{ij,1} - Y_{ij,0} | D_{ij,1} = 0] = ATT(1)
\end{equation} 
The equality holds only for $t < 2$, that is, for $t = 1$, as we are able to identify the average treatment effect for $t=1$ for treatment units starting treatment in period $t=1$.

If assumptions (i)--(iii) hold, then the OLS coefficient $\hat{\delta_{1}}$ in \eqref{did_model} is a consistent estimator of $ATT(1)$, the average treatment effect on the treated at period 1. The average treatment effect on the second cohort of treated units, $\delta_{2}$, is not identified as there are no longer any untreated units available for constructing counterfactuals. However, the OLS estimate of $\delta_{2}$ can inform us of the influence of initial exposure to the treatment if the effect for those longer exposed to treatment is the same for those just starting treatment. We cannot talk about the \emph{dynamic heterogeneity} of treatment when interpreting $\delta_{2}$, as we are making a not allowed comparison between two treated groups, we can just infer the difference in outcomes between those in the first and the second treatment cohort in the last treatment period.

\section{Data}\label{data}

For our main analysis, we collected Italian tweets posted before (up to 53 days before the lockdown) and after (14 days into the lockdown) the lockdown in Codogno. We collected Twitter data using the official Twitter Stream API:\footnote{See, \url{https://developer.twitter.com/en/docs/tutorials/stream-tweets-in-real-time}.} data collection started on January 1, 2020 and ended on March 22, 2020. When tweets are downloaded via API the individual JSON fragments representing a single tweet (the Twitter API returns tweets as JSON data structures) are associated with a user name, information regarding the user's activity (number of friends, followers, total count of tweets,  date of creation of the account, and others) and a user-defined location. 

Social media-based measures of uncertainty and sentiment have been proven to be a powerful tool that accurately reveals information about the emotional well-being of people during the COVID-19 pandemic \autocite{Metzler, 9207881}. Measures of uncertainty based on Twitter data proved to correlate with other uncertainty indicators computed on data from newspapers, stock market volatility, or forecaster disagreement, strengthening the validity of Twitter-based indicators of emotions \autocite{Altigetal2020}. Online and social media-based measures are powerful tools to accurately identify uncertainty and negative sentiments following major events, such as the COVID-19 pandemic \autocite{ 10.1057/s41599-022-01181-w, https://doi.org/10.48550/arxiv.2005.11458}.

In order to establish the quasi-experimental approach, we designated the municipalities that were subjected to the lockdown imposed by the Italian Government as the treated group. These municipalities were considered to be at the epicenter of the COVID-19 outbreak---i.e., the `red zone'. Conversely, we selected the non-urban, surrounding area of Codogno, identified by the Italian government as the `orange zone,' as our control group.\footnote{See, \textit{Gazzetta Ufficiale} n.45 (Feb. 23, 2020):\url{https://www.gazzettaufficiale.it/eli/id/2020/03/09/20A01521/sg}. To ensure the highest degree of similarity and comparability, we opted to select only the closest non-urban areas to the red zone of Codogno. The orange zone, as stated in the  Decree \autocite{dpcm1}, is broader and includes also the Milan urban area.}  

There are two indicators that can be used to assign the correct location of Twitter users: the user's self-reported location and the geo-tagged location of the tweet.\footnote{For a review of location-based methods, see \cite{Zheng2018}. See \cite{DEROSIS2021} for an application related to the COVID-19 pandemic.} The self-reported location allows fine-tuning the user's location without the need to count only geo-tagged tweets where the user is using his GPS to show its exact position. Because of the very limited geographic area covered by the analysis, GPS data may not accurately distinguish people inside and outside the red zone. User location is available and accurate for a fraction of 20-30 \% of all users (extensive tests were conducted in \autocite{DEROSIS2021} with a much larger sample). Conversely, the remaining part of Twitter users can report no position or use a fictional one.\footnote{It is worth noticing that machine learning techniques are able to infer their location at least at the regional level \autocite{Zheng2018}.} 

To make sure tweets are from the Codogno municipality and surroundings, we used two filters: selecting ``Italian language" and ``geographical filtering" (to include just the municipalities in the red and orange zones). The latter allows us to retrieve tweets with GSP coordinates within the selected area of interest. Then, we extract and clean the self-reported location of each tweet in order to correctly identify the municipality of origin of the tweets and control for the possible inaccuracies of locations retrieved solely via GPS.\footnote{Manual checks on self-reported location of tweets also were carried out to check whether the user location was coherent with the contents of the tweets. For example, we looked for tweets discussing life in isolation and confirmed that these were Twitter from the lockdown area.} 

\begin{figure}[H]
\centering
    \includegraphics[height = 0.6\textwidth]{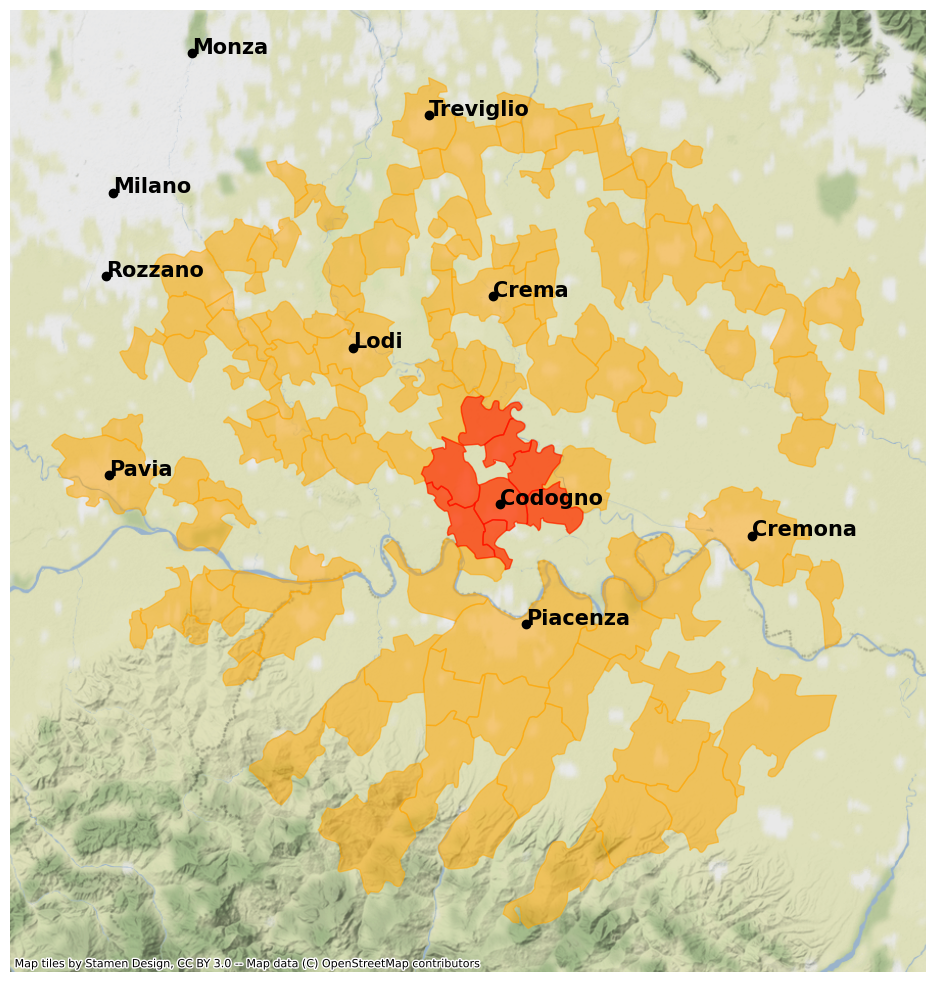}
  \caption{\emph{Red} zone and \emph{orange} zone municipalities featured in the analysis.}
  \label{fig:map_zones}
\end{figure}

After dropping duplicated tweets and re-tweets, we obtain a sample of 33,317 unique tweets. Of these, we obtain 2,005 tweets from the red zone and 31,312 from the orange zone cities. The tweets pertain to 1,277 Twitter accounts from the red and orange zone. The red zone covers eight of the ten lockdown locations in Lombardy (Casalpusterlengo, Codogno, Maleo, Somaglia, Fombio, Bertonico, Castelgerundo, Castiglione D'Adda). We observe no tweets from the small areas of San Fiorano (1\,849 inhabitants) and Terranova dei Passerini (731 inhabitants), as no user location matching the two municipalities is recovered in our sample. The orange zone encloses a total of 125 unique locations. Figure \ref{fig:map_zones} highlights the cities with at least one active georeferenced user account featured in our analysis.

Additionally, we collected a large sample of tweets from all over Italy (774\,407 tweets after removing duplicates and re-tweets) posted within the same time span mentioned above, by resorting to language filtering and word-based queries of coronavirus-related terms, such as ``Covid'',  ``Coronavirus''. We used these tweets for our placebo analyses detailed in Section \ref{placebo_sect}.\footnote{Using geographical coordinates of municipalities retrieved via ISTAT (year 2020) we are able to identify additional unique tweets: one new unique tweet from the red zone located in Castelgerundo, 109 from 27 new orange zone municipalities in close proximity to Codogno (Cavenago D'Adda, Monticelli D'Ongina, Borghetto Lodigiano, Castelsangiovanni, Turano Lodigiano) and 1,571 (167) from orange (red) zone municipalities already observed in the sample collected via geographical filtering. The new observations are added to the sample of tweets collected around Codogno.} 

Some control/not-yet-treated units enter lockdown one day before the nationwide extension of the measures on March 9, 2020. Because estimation on the treatment effect for these units using as counterfactual the rest of yet-to-be treated units will be very imprecise, we drop all tweets from March 7, 2020---the day of the announcement of the first extension of the quarantine rule---and from March 8, 2020. Also, to avoid any potential anticipation of treatment around the time of the discovery of the first COVID-19 transmission case in Codogno (February 20, 2020), we drop observations spanning from February, 20 to February 22, 2020. We are left with a final sample of 28,370 unique tweets, 1,604 from the red zone and 26,766 from the orange zone, accounting for 61 and 1,022 unique user accounts respectively.

\section{The effects of lockdown measures on uncertainty and sentiments}\label{results}

We use the collected data to estimate the coefficients of equation \ref{did_model} for general public response and across different topics of interest for the population and policymakers. In Section \ref{sec:identification} we discuss the identification of our model, while in Section \ref{subsection:results} we depict the results.

\subsection{Identification}\label{sec:identification}

We exploit the exogenous shock of the unexpected lockdown measures enacted in the red zone on February 23, 2020, to identify the impact of lockdown restrictions on public emotions. 

As explained in detail in Section \ref{sec:random}, the discovery of the first case of contagion in Codogno was random since the red and the orange zone were exposed to the same transmission potential of COVID-19 and the virus was already spreading in the orange zone before the extension of the lockdown measures \autocite{CEREDA2021100528}. The only difference between the red zone and the orange zone, which are also very close geographically, is that only the former was subject to the first lockdown.
Hence, individuals in the orange zone represent an ideal control group for the effect of the treatment lockdown policy on individuals in the red zone, who were subject to treatment. Furthermore, the lockdown restrictions of February 23, 2020, could not have been strategically anticipated, ruling out any possible bias deriving from manipulation of the treatment. 

Moreover, as shown in Section \ref{app:random} of the Appendix, the cities in the treated and control areas shared homogeneous socio-economic and demographic characteristics before the implementation of the lockdown measures. Covariance balance can serve as evidence that the treatment assignment approximates one of a randomized experiment, showing that the public reactions within the two groups did not systematically differ before the treatment assignment \autocite{loveplot}. The covariate balance check that we run does not identify any systematic difference between the two groups before the assignment to treatment (see Section \ref{test_assumptions}).

The present design makes a compelling case for the causal identification of the treatment effect in the present design, as the selected cities in the orange zone provide a suitable counterfactual of the area affected by the treatment or the lockdown policy. Our quasi-experimental setting allows us to reasonably argue for the existence of pre-treatment common trends among our control and treated units.

\subsection{Results}\label{subsection:results}

Table 1 reports the estimated effect of the COVID-19 shock on uncertainty and negative sentiments, aggregated (columns (1) and (6)) and split by topics---economics (columns (2) and (7)), health (columns (3) and (8)), politics (columns (4) and (9)), and lockdown policy (columns (5) and (10)). We cluster standard errors at the city level (119 clusters) using the Liang-Zeger formula \autocite{10.1093/biomet/73.1.13}.\footnote{The model estimates with White standard errors are provided in the Appendix, Section \ref{white_ses}.} With clusters, the DiD estimator is asymptotically normally distributed as the number of clusters grows large \autocite{roth2023whats}. Given that the number of treated clusters in our setting is relatively small, it is difficult to assume that there is a super-population of treated clusters from which we are independently sampling. Following \textcite{Rambachan2020DesignBasedUF}, clustering is appropriate---although potentially conservative---at the level at which we have independent treatment assignment. In our setting, clustering at the municipality level is a suitable choice as the treatment is assigned to municipalities independently, whereas units within the same cluster have perfectly correlated treatment assignments and are either all treated or not treated. 

\begin{table}[H]\centering
\def\sym#1{\ifmmode^{#1}\else\(^{#1}\)\fi}
\caption{DiD Regression table for \emph{Uncertainty} and \emph{Negative Sentiment}, aggregated and grouped by topics with user fixed effects.\label{did_1}}
\begin{adjustbox}{max width=\textwidth}
\begin{tabular}{lrrrrrrrrrr}
\toprule
&\multicolumn{5}{c}{Uncertainty}&\multicolumn{5}{c}{Negative Sentiment}\\
\toprule
                    &\multicolumn{1}{c}{(1)}&\multicolumn{1}{c}{(2)}&\multicolumn{1}{c}{(3)}&\multicolumn{1}{c}{(4)}&\multicolumn{1}{c}{(5)}&\multicolumn{1}{c}{(6)}&\multicolumn{1}{c}{(7)}&\multicolumn{1}{c}{(8)}&\multicolumn{1}{c}{(9)}&\multicolumn{1}{c}{(10)}\\
                    &\multicolumn{1}{c}{Aggregate}&\multicolumn{1}{c}{Economics}&\multicolumn{1}{c}{Health}&\multicolumn{1}{c}{Politics}&\multicolumn{1}{c}{Policy}&\multicolumn{1}{c}{Aggregate}&\multicolumn{1}{c}{Economics}&\multicolumn{1}{c}{Health}&\multicolumn{1}{c}{Politics}&\multicolumn{1}{c}{Policy}\\
\midrule
post=1              &      0.0378\sym{**} &     0.00819\sym{*}  &      0.0592\sym{***}&    -0.00679         &      0.0198\sym{***}&     -0.0337         &   -0.000346         &      0.0365\sym{***}&     -0.0108         &     0.00736\sym{***}\\
                    &      (2.94)         &      (2.52)         &      (4.94)         &     (-1.81)         &      (5.79)         &     (-1.76)         &     (-0.07)         &      (4.58)         &     (-1.79)         &      (3.99)         \\
\addlinespace
post=2              &      0.0500\sym{**} &      0.0169\sym{***}&      0.0673\sym{***}&    -0.00253         &      0.0191\sym{***}&     -0.0405\sym{*}  &     0.00614         &      0.0401\sym{***}&     -0.0159\sym{*}  &     0.00783\sym{***}\\
                    &      (2.96)         &      (3.47)         &      (6.04)         &     (-0.62)         &      (3.55)         &     (-2.12)         &      (1.12)         &      (5.20)         &     (-2.51)         &      (4.02)         \\
\addlinespace
red zone=1 $\times$ post=1&       0.145\sym{*}  &     0.00222         &      0.0606\sym{***}&      0.0158\sym{***}&      0.0281\sym{**} &     -0.0168         &     -0.0176         &      0.0614\sym{***}&      0.0281\sym{***}&      0.0389\sym{*}  \\
                    &      (2.37)         &      (0.34)         &      (4.48)         &      (3.40)         &      (3.01)         &     (-0.48)         &     (-1.49)         &      (3.46)         &      (3.66)         &      (2.49)         \\
\addlinespace
red zone=1 $\times$ post=2&      0.0475         &    -0.00495         &     0.00294         &    -0.00647         &   -0.000174         &     -0.0224         &     -0.0115         &      0.0419         &      0.0358\sym{***}&      0.0233         \\
                    &      (0.69)         &     (-0.54)         &      (0.15)         &     (-1.07)         &     (-0.01)         &     (-0.53)         &     (-1.39)         &      (1.85)         &      (3.57)         &      (1.70)         \\
\addlinespace
Constant            &       0.902\sym{***}&     -0.0119         &     -0.0702\sym{***}&     0.00900\sym{*}  &     -0.0189         &      0.0629         &     0.00535         &     -0.0819\sym{***}&     -0.0199\sym{*}  &     -0.0311\sym{*}  \\
                    &     (13.45)         &     (-1.54)         &     (-4.56)         &      (2.02)         &     (-1.57)         &      (1.65)         &      (0.87)         &     (-3.84)         &     (-2.57)         &     (-2.29)         \\
\midrule
Observations        &       28370         &       28370         &       28370         &       28370         &       28370         &       28370         &       28370         &       28370         &       28370         &       28370         \\
Clustered SE & Yes & Yes & Yes & Yes & Yes & Yes & Yes & Yes & Yes & Yes \\ 
\bottomrule
\multicolumn{11}{l}{\footnotesize \textit{t} statistics in parentheses}\\
\multicolumn{11}{l}{\footnotesize \sym{*} \(p<0.05\), \sym{**} \(p<0.01\), \sym{***} \(p<0.001\)}\\
\end{tabular}
\end{adjustbox}
\end{table}

We first look at the effect of the lockdown on aggregate uncertainty and negative sentiments. In Column (1), the estimates show that the probability of expressing uncertainty increases over time for both the treated and control groups, as the estimated coefficients of $post=1$ (0.0378) and $post=2$ (0.500) are positive and statistically significant. The treatment effect ($red zone=1$ $\times$ $post=1$) is positive and statistically significant at 0.05. 

Column 6 indicates that the lockdown does decrease the general negative sentiment, but also that it does not significantly impact the reaction of those in the \emph{red zone}. Overall, the aggregate results indicate that the lockdown had no effect on aggregate uncertainty and negative sentiments for those subject to it. However, as we motivated, the lockdown may have impacted heterogeneously different types of uncertainty and sentiments. For this reason, it is worth exploring the disaggregation of uncertainty and sentiments by topic.  

The impact of the lockdown on economic uncertainty and sentiments is displayed in columns (2) and (7) respectively. Column (2) reports the DiD regression estimates on economic uncertainty. The significant and positive estimated coefficients of the shared time trend ($post=1$ and $post=2$) indicate that economic uncertainty increases as lockdown measures expand geographically. This result is in line with previous findings that identified a spike in economic uncertainty when COVID-19 hit \autocite{Baker2020}. However, the lockdown does not significantly affect the economic uncertainty for those under the lockdown. Column (7) depicts the effects of negative sentiments on economic matters. Again, the lockdown does not hold any significant impact on economic sentiment. These results show that the lockdown had no major impact on economic uncertainty and negative sentiments. 

Turning to health-related matters, in Column (3) we see a significantly increasing common trend in health-related uncertainty following the start of the local lockdown (0.0592) and the later nationwide expansion of the policy (0.0673). The results show that the lockdown had a significant and positive effect on health-related uncertainty, as indicated by the coefficient of the interaction. This effect disappears once the policy is applied nationwide. Negative sentiments in health-related tweets, shown in Column (8), present similar results. However, differently from the other significant results, health-related negative sentiments violate the parallel trend assumption, as we explain in the next section. The lockdown increased negative sentiments among the treated (0.614). The general trend shows a significant deterioration of sentiments around health issues, as indicated by the coefficients of $post=1$ and $post=2$ (0.0365 and 0.0401 respectively). 

The results related to the political uncertainty and sentiments are in Columns (4) and (9) respectively. In terms of uncertainty, the lockdown has a significant impact on political uncertainty, which increases by 0.0158 among users in the treated area. Interestingly, the same direction of the results is depicted for negative sentiments. Column (9) displays a negative and significant coefficient for the $red zone=1\times post=1$ interaction. This indicates that the lockdown sentiment around politics had an improving trend which was the opposite among the treated units for which the negative sentiment significantly increased after the lockdown was put in place (0.0281). This result, which holds also when the measures are extended at the national level ($red zone=1\times post=2$), partially contrasts with the literature that identified a positive impact of lockdown policies on public opinion toward the incumbents (e.g., \autocite{BOL2020}).  

Column (5) displays the results for uncertainty related to the policy itself. Although the coefficients of $post=1$ and $post=2$ show that the common trend uncertainty around the practical implications of the lockdown is significantly above baseline (0.0198 and 0.00191 respectively), being subject to the lockdown does significantly increase the worries of those subjected to the measures. Column (10) displays a similar trend for sentiments: negative sentiment around the policy increases everywhere and particularly in the lockdown areas.

Our estimates show that the lockdown impacted emotions related to health, politics, and the lockdown policy, but had no major consequences for uncertainty and sentiments related to economics. These results are particularly interesting as they allow us to disentangle the actual effect of the lockdown policy from the general trends in emotions driven by the shock of the pandemic. This comparison is possible by comparing the coefficients of the interaction with those of the time-dependent dummies, that display the trends in emotions in the control group. Applying this distinction is generally very difficult, but can be achieved in a quasi-experimental setting like the one we provide in this paper. 

Based on these results, we draw three main takeaways. First, the growth in economic uncertainty is purely COVID-driven. By contrast, the lockdown does not have a significant impact on economic sentiments. This finding is key as it shows that implementing a lockdown does not come at the costs of higher economic uncertainty and worsened economic sentiments, which have repercussions on financial and economic variables \autocite{Baker2016, Adamsetal2023}.

Second, being subject to a lockdown makes people more concerned about health issues. This finding is not trivial. Lockdowns can signal the intention of the public authority to tackle the virus with all means. Moreover, the enforcement of a lockdown should shape expectations positively, as people should expect COVID-19 cases to reduce once strict rules prevent the diffusion of the virus. If this was true, we would observe less distress in lockdown areas compared to the untreated areas---where our estimates show, indeed, that health uncertainty and sentiments are worsening. We find that this is not the case. Our results indicate that the policy increased uncertainty and negative sentiments among individuals in the treated area. This means that the lockdown has exacerbated the already growing concerns. It is worth noticing that higher uncertainty and negative sentiments on health issues may not necessarily represent a drawback of the policy (while this would have been the case for economic issues, where the negative effect of emotions on the economy is incontrovertible). It is true that, on the one hand, negative emotions about health issues reflect people's anxiety over their safety, which is associated with negative repercussions on their mental health. However, on the other hand, heightened concerns around health may be beneficial, as they increase public awareness around the health risk posed by the virus. For this reason, the worsening of health sentiments should not necessarily be seen as a cost of the lockdown measure, as it could have some benefits too. 

Third, the lockdown worsened public sentiments toward politics. This result is particularly surprising in light of the recent literature that analyzed the extraordinary changes in public opinion toward politicians during COVID-19. Previous work found that attitudes toward politicians improved during the pandemic \autocite{BOL2020, DeVries2021, Sibley2020}. This led some to advance the hypothesis that lockdown measures may have contributed to boosting such support \autocite{BOL2020}. We find that this is not the case. Indeed, we show that sentiment around politics worsens significantly among users in the treated area. We verify that such worsening is potentially led by uncertainty and sentiments around the policy itself (Columns (5) and (10) of Table \ref{did_1}). This result indicates that lockdown policies come at the major cost of worsened political emotions.

\subsection{Robustness checks}\label{robustness_checks}

We perform a battery of tests to assess the robustness of our results. Here we discuss the main takeaways.

First, we assess the robustness of our results with respect to the assumptions of the DiD model. We start by performing statistical tests of the hypothesis of the differences in pre-treatment trends (Section \ref{sec: pre-trends}) to look for possible violations of the \emph{parallel trends} assumption of the DiD model -- i.e., that the difference between the treated and control groups is constant before the implementation of the treatment \autocite{kahn-lang}. We test for differences in pre-treatment trends between treated and control groups by regressing the DID model on \emph{m} leads and \emph{q} lags of the treatment variable over several time periods, choosing as a baseline period the three weeks before the start of the treatment. The pre-treatment trends in  both uncertainty and negative sentiment about politics and the policy itself are not statistically different between the treated and controls, which is evidence in favor of the parallel trend assumption. The same applies to health uncertainty. Instead, pretreatment trends in aggregate uncertainty and negative sentiment about health differ substantially between the two areas, suggesting a violation of the parallel trends condition. 

We also allow for violations of the parallel trends conditions following \textcite{RambachanRoth23} (Section \ref{honestdid_sect}), by assuming that the magnitude of the violation of parallel trends after treatment cannot be greater than $\bar{M}$ times the largest violation before treatment. We can allow for different post-treatment violations of parallel trends and still retain a robust significant effect for uncertainty and negative sentiments towards health and the policy and negative sentiments towards politics. The impact of the lockdown on uncertainty related to politics is not robust to any violations in parallel trends, suggesting that the lockdown is significantly affecting the concerns around politics only if we allow for zero violations, as required by the usual specification of DiD model assumptions.

Second, we jointly test the p-values of the models using the Benjamini-Hochberg adjusted p-values \autocite{BH} for testing multiple hypotheses and show that the original small p-values of the effect of the lockdown hold even when we adjust for correlations between the regression models that might lead us to incorrect conclusions. Detailed results can be found in Section \ref{adj_pval_sect} of the Appendix.

Finally, in Section \ref{placebo_sect} of the Appendix, we conduct a \emph{placebo test} on the effects of the lockdown when the lockdown is not unexpected---i.e. considering the effects of the national lockdown on public reaction. We focus on northern Italy and take as \emph{placebo treated group} all tweets originating from cities with monthly excess mortality in January and February 2020 \emph{closest} to the red zone average in terms of Euclidean distance. As \emph{placebo controls} we take the tweets from cities in the north with monthly excess mortality \emph{most distant} from the red zone. Because the national lockdown has no significant effect on the public reaction of the placebo treatment group, our results are robust to the anticipatory effects of the treatment.
  
Overall, our robustness tests corroborate some common results.
The lockdown effect passes all stress tests and indeed has a significant effect on uncertainty around health, politics, and  the policy itself, as well as on negative sentiments surrounding the policy and politics. The effect on uncertainty towards politics is robust only if we assume that the parallel trends condition is not violated, which is consistent with the basic setup of a DiD model. Thus, under the assumption of parallel trends without violations, the effect is robust for most of our results, except for negative sentiments related to health and aggregate uncertainty, because we rejected the hypothesis that there was no difference in trends before treatment, making it hard to justify the assumption of parallel trends in those cases.

\section{Conclusions} \label{conclusion}

A major concern for policymakers when implementing lockdown measures is to balance the health benefits against the negative socioeconomic consequences, including heightened uncertainty and negative sentiments. The spread of COVID-19 and the measures taken to contain infection evolved in parallel with rising uncertainty and negative emotions, which are associated with severe economic and social costs, posing a serious challenge to policymakers who need to assess the pros and cons of imposing such strict measures. It also poses a problem to scholars, who are faced with the challenge of disentangling the effect of the restrictions from that of the pandemic itself, which had a major impact on people's emotions and behaviors. This problem calls for a thorough analysis that identifies the causal impact of lockdown measures.

Machine learning techniques, combined with quasi-experimental econometric methods, can help us to provide a clearer picture of this research question. The quasi-experimental design allowed us to estimate the causal impact of the lockdown, based on the case of the first high-income economies to ever implement a COVID-19 stay-at-home mandate. Machine learning techniques applied to textual data from Twitter enabled us to observe and study the heterogeneous change in public emotions and how they shifted based on different topics. The approach we propose in this paper provides a more nuanced definition of uncertainty, which so far in the literature has tended to focus on economic uncertainty, that could be easily applied in other policy evaluation exercises.  

Overall, we find that indeed lockdowns come at a cost, which is however somewhat unexpected. The major cost we identify is not in terms of economic uncertainty and sentiments, where we find no causal impact of the policy, suggesting that economic-related sentiments may be mostly driven by the rise in COVID-19 cases rather than the policy. By contrast, we find that in the red zone, however, users were more likely to express uncertainty around health and politics and negative sentiments on politics. This evidence highlights that lockdown increase, rather than decrease, uncertainty around health conditions among users, which however do not seem to express sentiments that are significantly more negative.

Nevertheless, our results also indicate that the lockdown comes with non-negligible political costs, as the treated area experienced a surge in negative sentiments and uncertainty related to politics. This is an important concern for policymakers, as it could inhibit future timely implementations of stay-at-home mandates to avoid political backlash. In addition, the lockdown area expressed significantly higher uncertainty and discontent when discussing the policy intervention. This suggests that the behavioral guidelines associated with the lockdown measures were met with a high degree of confusion and may have not been effectively communicated to those living under the restrictions, leading to resentment towards the measure among those affected. Increasing health and policy concerns suggest that policymakers failed to increase public support for the health authorities and the health systems during the emergency.

Crucially, while our findings identify some costs associated with lockdown measures, they do not assess their efficacy in containing the virus, which has proven to be successful in the epidemiology literature. The evidence in this paper should not inhibit authorities from imposing lockdowns but, if anything, it should be used to ameliorate such policies in the event of future emergencies. Based on these results, policymakers could reduce the costs of rising uncertainty and negative sentiments by increasing the institutional efforts to appropriately communicate lockdown rules and the consequences of the restrictions to the population. 
By actively monitoring and responding to the public's needs, authorities could actively provide relief and reduce uncertainty and social panic by keeping the public informed with official and timely updates on social media. This would not only help alleviate public frustration but could also be used as an opportunity to build general support for public health authorities and the government's handling of the extraordinary situation.


\printbibliography

\appendix

\appendix
\begin{center}
    \textbf{\huge Appendix}
\end{center}

\vspace{0.5cm}

\section{Classification Model}\label{cls_model}

We fine-tune the pre-trained AlBERTo model \autocite{polignano2019alberto} on downstream tasks. 
The model's architecture is based on BERT \autocite{devlin2018bert}, which innovates the previous state-of-the-art based on unidirectional, left-to-right Transformer architectures by introducing a \emph{masked learning model} pre-training objective, which enables the model to understand left-to-right and right-to-left context of terms in a given text.


In its basic version, the one we employ with AlBERTo, BERT consists of of 12 encoders blocks, given by 12 layers, 12 \emph{self-Attention} heads and 768 hidden units. Each encoder consists of a self-attention layer and feed-forward layer - each sub-layer of the encoders being followed by a residual connection and a normalization layer. Attention mechanism is the building block of any Transfomer architecture. The self-attention layer will improve the encoding of a given input token by looking at all other words in the input sequence in order to convey information on the influence of the context. 

We initialize the pre-trained (base) AlBERTo by loading the HuggingFace Transformers model made available by \cite{polignano2019alberto}\footnote{\url{https://github.com/marcopoli/AlBERTo-it}}. The model is pre-trained on the TWITA dataset, a large corpus of 200 mln tweets written in Italian from 2012 to 2015 . 

Normalization of tweets is achived via Ekphrasis \autocite{baziotis-pelekis-doulkeridis:2017:SemEval2}. Sentences are normalized to tag urls, numbers, hashtags, emoticons. All punctuation is removed, except emojis, question and exclamation marks and accented characters. The text is then tokenized with SentencePiece \footnote{\url{https://github.com/google/sentencepiece}}, a sub-word segmentation algorithm that probabilistically samples and tries out multiple segmentations of the inputs via sub-word regularization.  

In order to fine-tune AlBERTo on our specific classification tasks, we first manually assign $\sim 20\%$ of the tweets (6,318) to possible levels of \emph{Uncertainty} and \emph{Sentiment}: \emph{low}, \emph{medium} or \emph{high}. The classes we obtain are highly imbalanced, as the majority of tweets end up belonging in the \emph{medium} class. We proceed with the application of AlBERTo in a two-step classification process. \emph{Step 1} classifies tweets as \emph{medium (1) vs rest (0)}, \emph{Step 2} will assign the non-medium tweets of step 1 to the remaining categories \emph{low} (0) vs \emph{high} (1), corresponding to \emph{uncertainty} (0) vs \emph{certainty} (1) or \emph{negative} (0) vs \emph{positive} (1) sentiment. Therefore, we fine-tune AlBERTo model separately for each step of the two tasks of detecting uncertainty and negative sentiments, for a total of four fine-tuned models. 

As indicated by \cite{https://doi.org/10.48550/arxiv.1810.04805}, small datasets are somewhat sensible to hyperparameter choice when fine-tuning on BERT. Therefore, we run a hyperparameter grid-search with Optuna \footnote{\url{https://github.com/optuna/optuna} \autocite{akiba2019optuna}} to select the best performing fine-tuned model. Starting from the suggestions in the BERT original paper, we search over the following set of values:
\begin{itemize}
    \item Batch size: $[16, 32]$
    \item Seed (random initialization for new parameters W) :  $[40,41,42]$
    \item Learning rate (LR): $[5e-5, 3e-5, 2e-5]$
    \item Warmup steps:  $[0.001, 0.01, 0.1]$
    \item Weight decay:  $[100, 1000, 10000]$
\end{itemize}
The grid-search is done separately for each one of the four classification tasks. We split the manually annotated dataset into \emph{train}-\emph{validation}-\emph{test} sub-samples, with 50-30-20 proportions. The model is trained on the training sample and evaluated at the validation sample. We let the training run for up to 15 epochs, implementing early stopping on validation loss to avoid overfitting with patience equal to 3 epochs. For the Adam optimization algorithm and the dropout probability, parameters remain the same of the pre-training ($\beta_{1}=0.9$, $\beta_{2}=0.999$ and Dropout = 0.1). 
The grid-search selects the following models as the best in terms of decrease in validation loss at early stopping:
\begin{itemize}
    \item Uncertainty:
    \begin{enumerate}
        \item Step 1 - medium vs rest: batch size = 16, seed = 41, LR = 5e-05, weight decay = 0.1, warmup steps = 10000. Best model at 3 epochs. 
        \item Step 2 - Uncertainty vs Certainty: batch size = 32, seed = 41, LR = 2e-05, weight decay = 0.1, warmup steps = 10000. Best model at 15 epochs. 
    \end{enumerate}
        \item Negative Sentiment:
    \begin{enumerate}
              \item Step 1 - medium vs rest: batch size = 16, seed = 42, LR = 2e-05, weight decay = 0.001, warmup steps = 10000. Best model at 4 epochs. 
        \item Step 2 - Negative vs Positive Sentiment: batch size = 32, seed = 40, LR = 3e-05, weight  decay = 0.1, warmup steps = 10000. Best model at 13 epochs. 
    \end{enumerate}
\end{itemize}

We set the hyperparameters at the values selected by the grid-search and proceed to fine-tune the pre-trained AlBERTo on each classification task, using the training and validation subsamples of manually-classified tweets.
We then use the fine-tuned model to classify the rest of tweets without manual annotations.
For each step of classification, we report the precision, recall, F1-Score of each class and both the macro and weighed average values (Table \ref{perf_1}) obtained by using the fine-tuned models to predict labels from the test sample. In Table \ref{perf_2}, macro average accuracy and average precision scores are displayed, as well as the macro ROC-AUC. All classification performance measures were computed in Python using SkLearn API \autocite{sklearn_api}.

Fine-tuning the Alberto model directly on a multi-label classification task performs far worse than the two steps procedure, which performs far better. The grid search on the multi-label AlBERTo reveals a much lower performance shared across different combinations of hyperparameters, achieving around 60\% accuracy on the validation sample. Instead, in the two-steps procedure for binary classification, AlBERTo is able to efficiently distinguish between the medium class and the low-high class.
 

\begin{table}[H]\centering
\def\sym#1{\ifmmode^{#1}\else\(^{#1}\)\fi}
\caption{Classification Report for fine-tuning of pre-trained Alberto on Uncertainty and Negative Sentiment classification tasks (step 1 and step 2). \label{perf_1}}
\begin{adjustbox}{max width=\textwidth}
\begin{tabular}{lrrrrr}
\toprule
\multicolumn{2}{c}{\textbf{Uncertainty}} & Precision &    Recall &  F1-score &  Support\\
\midrule
\multirow{4}{*}{\textbf{Step 1}}
&\emph{Medium}&   0.726783 &  0.809595 &  0.765957 &   667 \\
&\emph{Rest}  &   0.756238 &  0.659966 &  0.704830 &   597 \\
& macro avg    &   0.741511 &  0.734781 &  0.735394 &  1264 \\
& weighted avg &   0.740695 &  0.738924 &  0.737086 &  1264 \\
\midrule
\multirow{4}{*}{\textbf{Step 2}} 
&\emph{Uncertainty}&   0.707463 &  0.707463 &  0.707463 &  335 \\
&\emph{Certainty} &   0.720000 &  0.720000 &  0.720000 &  350 \\
& macro avg    &   0.713731 &  0.713731 &  0.713731 &  685 \\
& weighted avg &   0.713869 &  0.713869 &  0.713869 &  685 \\
\toprule
\multicolumn{2}{c}{\textbf{Negative Sentiment}} & Precision &    Recall &  F1-score &      Support \\
\midrule
\multirow{4}{*}{\textbf{Step 1}} 
&\emph{Medium}&   0.645872 &  0.713996 &  0.678227 &   493 \\
&\emph{Rest} &   0.803894 &  0.749676 &  0.775839 &   771 \\
& macro avg    &   0.724883 &  0.731836 &  0.727033 &  1264 \\
& weighted avg &   0.742260 &  0.735759 &  0.737767 &  1264 \\
\midrule
\multirow{4}{*}{\textbf{Step 2}} 
&\emph{Negative Sentiment} &   0.856734 &  0.911585 &  0.883309 &  328 \\
&\emph{Positive Sentiment}  &   0.806667 &  0.707602 &  0.753894 &  171 \\
& macro avg    &   0.831700 &  0.809594 &  0.818601 &  499 \\
& weighted avg &   0.839576 &  0.841683 &  0.838960 &  499 \\
\bottomrule
\end{tabular}
\end{adjustbox}
\end{table}

\begin{table}[H]\centering
\def\sym#1{\ifmmode^{#1}\else\(^{#1}\)\fi}
\caption{ROC AUC, Average Precision Score and Balanced Accuracy Score with fine-tuning of pre-trained Alberto on Uncertainty and Negative Sentiment classification tasks (step 1 and step 2). \label{perf_2}}
\begin{adjustbox}{max width=\textwidth}
\begin{tabular}{lrrrrr}
\toprule 
& Classification Step & Accuracy &    PR-AUC &   ROC-AUC  \\
\midrule
\multirow{2}{*}{\textbf{Uncertainty}} 
& Step 1 & 0.738924  &  0.659693 &                 0.734781 \\
& Step 2 & 0.713869 & 0.661466 &                 0.713731 \\
 \midrule
\multirow{2}{*}{\textbf{Negative Sentiment}} 
& Step 1 & 0.735759 & 0.75535 &                 0.731836 \\
& Step 2 &   0.841683 & 0.671 &                 0.809594 \\
\bottomrule
\end{tabular}
\end{adjustbox}
\end{table}

\section{Topic Dictionaries and WordClouds}\label{topics_appendix}

\begin{figure}[H]
  \begin{subfigure}[h]{0.5\textwidth}
    \includegraphics[width=\textwidth]{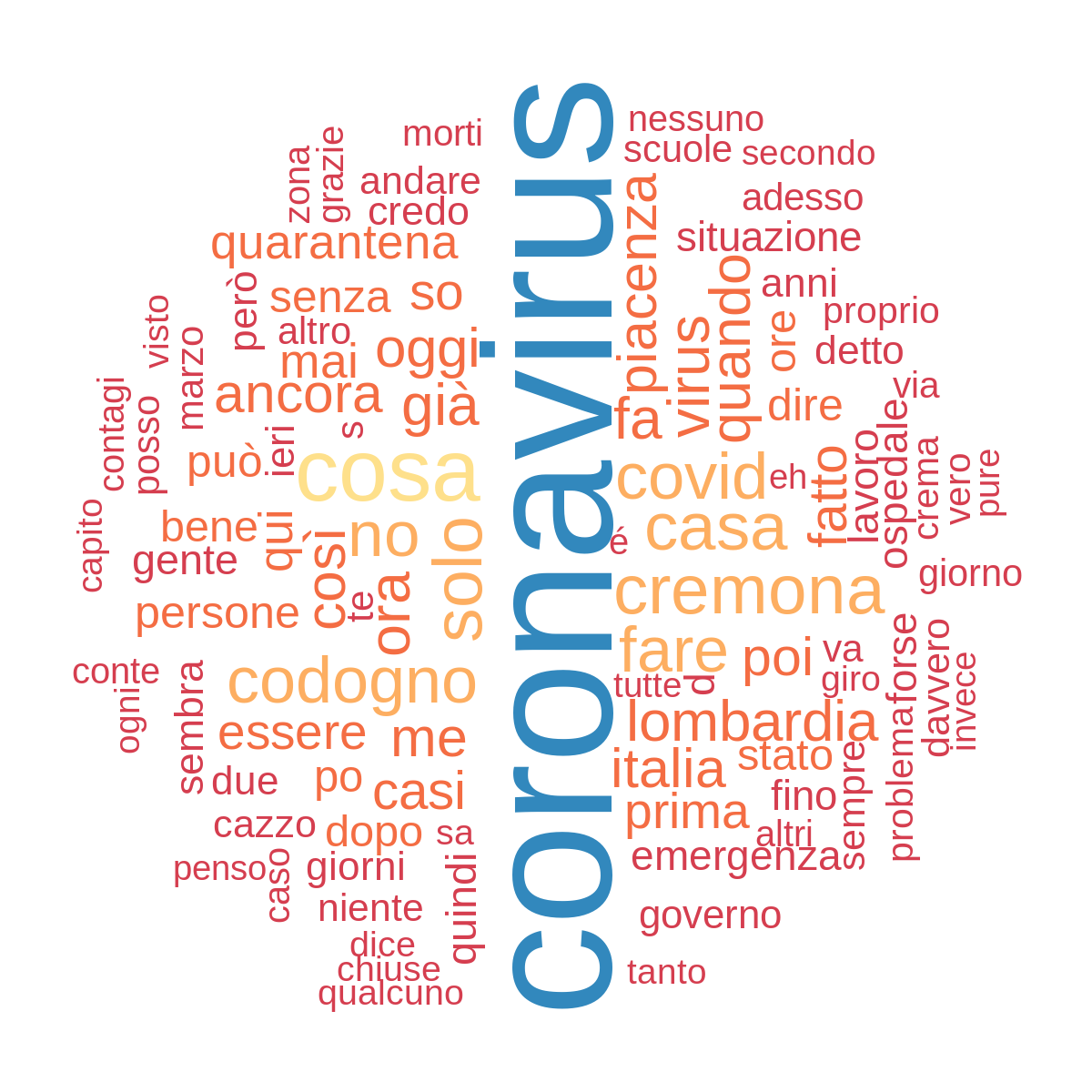}
    \caption{Uncertainty}
    \label{fig:f1}
  \end{subfigure}
  \hfill
  \begin{subfigure}[h]{0.5\textwidth}
    \includegraphics[width=\textwidth]{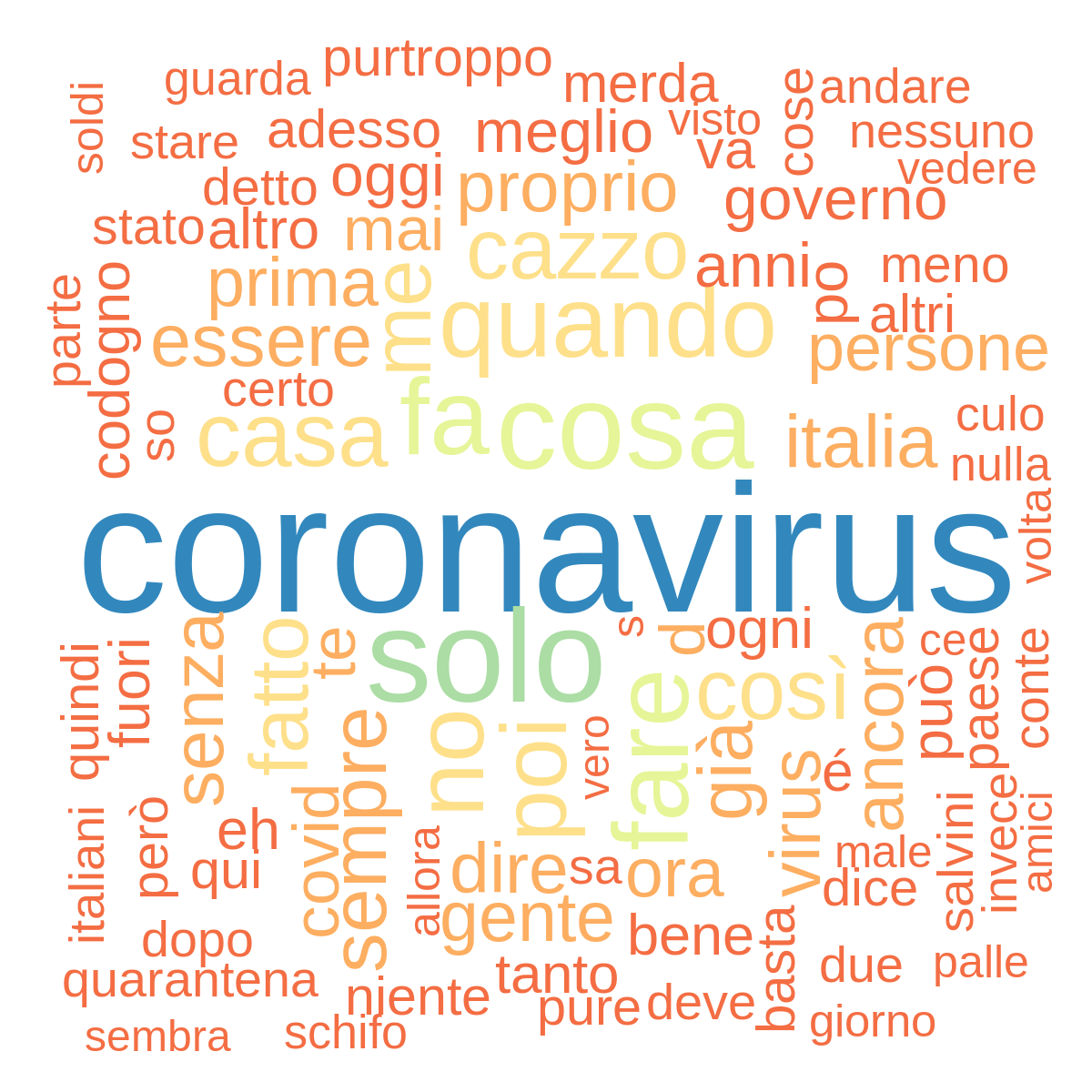}
    \caption{Negative Sentiment}
    \label{fig:f1}
  \end{subfigure}
  \caption{Wordclouds of first 100 most common terms features in tweets classified as \emph{Uncertainty} and \emph{Negative Sentiment}.}
  \label{dd2}
\end{figure}

\begin{table}[H]\centering
\def\sym#1{\ifmmode^{#1}\else\(^{#1}\)\fi}
\caption{Dictionaries for classification into \emph{Topics} in Italian \label{dictio_ita}}
\begin{center}
\begin{adjustbox}{max width=\textwidth}
\begin{tabular}{|c|}
\hline
\\
\multicolumn{1}{|c|}{\textbf{Economics}}\\
\hline
\\
economia, economic*, soldi, investiment*, banc*, finanz*, disoccupa*, bancarott*, imprenditor*, \\
impres*, lavoro, bonus, commerci*, gestione, piano, sostegno, crisi, iva, dipendent*, fiscal*, \\
coldiretti, agevolazioni, contribut*, reddito, salari*, confedilizia, confindustria,  cgil,\\
professionist*, negoz*, euro, mutuo, mutui, tasse, tassa, tassazione, tassat*, evasione, \\
fisco, sindacat*, inps, credit*, prestit*, stipendio, deficit,  lavorator*, produzion*,\\
produttiv*, aziend*, client*, soci, salone, ristorant*, smart-working, smart working, \\
commercial*, supermercat*, spes*, mercat*, turist*, turismo, licenzia* fiera, fiere, \\
cassa integrazione, lavorare, lavorare a casa. \\
\\
\hline
\\
\multicolumn{1}{|c|}{\textbf{Health}}\\
\hline
\\
coronavirus, virus, contag*, covid, tampon*, mascher*, casi, quaranten*, \\
mort*, ospedal*,
malat*, malatti*, sanità, sanitari*, medic*, medicina, infermier*,  
\\positiv*, farmaci*, kn95, terapi*, terapia intensiva, terapie intensive, sars, 
\\ sars-cov-2, paziente zero, pazient*, infett*, salute,decess*, influenza,
\\peste, sanita, guarit*, guarigion*,
ammarlarsi, ammalarci, ammalat*,\\
ammalare, covid19, croce rossa, epidemiolog*, oms, febbre, asintomatic*, \\ rianimazione, epidemia, respiratori, ricoverat*, portatore sano. \\
\\
\hline
\\
\multicolumn{1}{|c|}{\textbf{Politics}}\\
\hline
\\
politic*, govern*, italiaviva, salvini, renzi, conte, meloni, presidente, lega, ministro, \\
sindaco, decreto, legge, movimento5stelle, mattarella, segretari*, legislativo, parlament*,\\
giunta, assessor*, ue, politic*, profughi, pd, ong, sinistra, migranti, democrazia,\\
democratic*, partito, partiti, sardine, dimettiti, dimission*, fascismo, fascist*, \\ 5s, protesta, contedimettiti, nazismo, nazist*, destra, casta, m5s.\\
\\
\hline
\\
\multicolumn{1}{|c|}{\textbf{Policy}}\\
\hline
\\
chius*, sospes*, cancellat*, limitazion*, annullat*, chiud*, sospension*, isolat*, isolamento, \\
rinviat*, scorte, viveri necessari, zona rossa, zona arancione, luoghi di aggregazione, distanza, 
\\ restrizion*, controlli, posto di blocco, posti di blocco, spostamenti, autocertificazione.\\
\\
\hline
\end{tabular}
\end{adjustbox}
\end{center}
\end{table}

\begin{table}[H]\centering
\def\sym#1{\ifmmode^{#1}\else\(^{#1}\)\fi}
\caption{Dictionaries for classification into \emph{Topics} in English \label{dictio_eng}}
\begin{center}
\begin{adjustbox}{max width=\textwidth}
\begin{tabular}{|c|}
\hline
\\
\multicolumn{1}{|c|}{\textbf{Economics}}\\
\hline
\\
economy, economic*, money, investment*, bank*, financ*, unemployed, bankruptcy, entrepreneur*, \\
business*, work, bonus, trade, management, plan, support, crisis, VAT, employee*, fiscal*, \\
coldiretti, subsidies, contribution*, income, wage*, confedilizia, confindustria, cgil,\\
professional*, shop*, euro, mortgage, mortgages, taxes, tax, taxation, evasion, \\
taxman, trade union*, INPS, credit*, loan*, salary, deficit, worker*, production,\\
productive*, company*, client*, partners, salon, restaurant*, smart-working, smart working, \\
commercial*, supermarket*, expens*, market*, turist*, turism, licens*, fair, fairs, \\
layoffs, work, work at home. \\
\\
\hline
\\
\multicolumn{1}{|c|}{\textbf{Health}}\\
\hline
\\
coronavirus, virus, contagion*, covid, swab*, mask*, cases, quarantin*, \\
death*, hospital*,
sick*, sickness*, health, sanitary*, medic*, medicine, nurse*, 
\\positive*, drugs*, kn95, therapy*, intensive care, ICU, sars,
\\ sars-cov-2, patient zero, patient*, infected*, death*, flu,
\\plague, sanitation, heal*, ill*, recover*,\\
covid19, red cross, epidemiolog*, oms, fever, asymptomatic, \\ resuscitation, epidemic*, respiratory, hospitaliz*, carrier*. \\
\\
\hline
\\
\multicolumn{1}{|c|}{\textbf{Politics}}\\
\hline
\\
politic*, govern*, italiaviva, salvini, renzi, conte, meloni, president, lega, minister, \\
mayor, decree, law, 5stelle movement, mattarella, secretaries*, legislative, parliament*,\\
assessor*, eu, refugees, pd, ngo, left, migrants, democracy,\\ 
democratic*, party, parties, sardines, resign, resign*, fascism, fascist*, \\ 5s, protest, conteresign, nazism, nazist*, right, casts, m5s.\\
\\
\hline
\\
\multicolumn{1}{|c|}{\textbf{Policy}}\\
\hline
\\
clos*, suspend*, cancel*, restrict*, isolated, isolation, \\
postpon*, supplies, necessary food, red zone, orange zone, meeting places, distanc*,\\
controls, checkpoint, checkpoints, travel, self-certification.\\
\\
\hline
\end{tabular}
\end{adjustbox}
\end{center}
\end{table}

\section{Top fifty tweets}
\label{sec:top_tweets}

In this section, we provide some examples of tweets that presented the highest Shannon entropy scores for each emotion-topic pair. For each pair, we select five examples for illustration. We used Google Translate to translate each tweet from Italian to English and adjusted the translation whenever needed. We report in parenthesis and in bold the Shannon entropy score of each tweet. When necessary, we add some notes in square brackets to ease the interpretation of the text.

The classifier performs well in capturing sentiments, with a few minor drawbacks. The classifier proves successful in identifying the topics and detecting negative sentiments, even when the text is sarcastic. For instance, in the first tweet among health negative sentiments, the user sends `big congratulations' to the prime minister for his `perfect checks' on COVID-19 cases. Despite the use of these words, which out of context are considered positive, the sentiment of the tweet is identified as negative.
Negative sentiments in tweets encompass a range of negative emotions, including fear, anger, and anxiety. 

There is some inevitable overlap across topics, which however does not denote a mistake by the classifier, but stems from the actual overlap of topics in the text. For instance, some tweets with the highest entropy in terms of economic negative sentiments are also related to politics, as they blame `Italian politics' for the `incalculable economic damage' they created. The same happens for some politics-related tweets, where the Prime Minister is sarcastically praised for his `perfect checks' in light of the rising risks of coronavirus infections. Notably, this tweet appears in the top 50 of the categories `politics' and `health'. 

\subsection*{Economics Uncertainty}

\begin{itemize}
    \item It seems so. We await technical details on closing all commercial activities except for public utilities [\textbf{.99}]
    \item It will be the beginning of our financial monetary crisis that we will not be able to sustain which will force us to leave the URL [\textbf{.97}]
    \item \#fightcoronavirus hashtag \#cremona don't come to the bank better do everything online - appeal of the bank unions URL [\textbf{.93}]
    \item Very little money is saved and incalculable damage is done [\textbf{.91}]
    \item But are the migrant gentlemen in front of the supermarkets immune from the March decree or do they all have a self-certification? [\textbf{.75}]
\end{itemize}

\subsection*{Economics Negative Sentiments}

\begin{itemize}
    \item I live in \#Casalpusterlengo \#Iamblocked without being able to go to work. Wake up! [\textbf{.79}]
    \item The lightness of \#italianpolitics in facing \#covid is provoking incalculable economic damage [\textbf{.64}] 
    \item Meanwhile those who are part of the productive sector must not stay at home but go and get infected for the good of the capital [\textbf{.62}]
    \item \#Istayhome ah no I can't work in insurances or banks. \#factory i don't have the right to \#stayhome it's a shame [\textbf{.37}]
    \item you ruined italy: all the Italian sectors from the craftsman trader to tourism etc. [\textbf{.30}] 

\end{itemize}

\subsection*{Health Uncertainty}

\begin{itemize}
    \item I think that in November in the red zone there will be a demographic explosion of positive covid obviously [\textbf{.99}]

    \item Coronavirus Amendola [Minister of European Affairs]: it is possible that the EU [European Union] will give us more budget flexibility [\textbf{.99}]

    \item For the first time since the beginning of the emergency I have news of a person in intensive care and another who died [\textbf{.98}]

    \item By now we know who Salvini [leader of the right-wing political party] is. The problem is how much is true about this virus [\textbf{.98}]

    \item if we healthcare workers wore [face masks] as he did we would make a massacre [\textbf{.93}]
\end{itemize}

\subsection*{Health Negative Sentiments}

\begin{itemize}
    \item \#Conte [Italy's prime minister] \#wakeup the \#coronavirus advances and reached Lodi [one of the main cities in the red zone] big congratulations to you, perfect checks [\textbf{.93}]
    \item My mom is in the hospital and they don't have face masks \#coronavirus \#covid \#italy and they made everyone swab [\textbf{.90}]
    \item  more than a month to declare we are ready everything is under control and then the infection comes  [\textbf{.76}]
    \item number of infections from  \#coronavirus are on the rise but in \#Piacenza there is no outbreak [\textbf{.67}] 
    \item we laugh and joke we side between optimists and pessimists but patient zero [the first patient to be identified] is still around sowing [i.e. spreading the virus] [\textbf{.61}]

\end{itemize}

\subsection*{Politics Uncertainty}

\begin{itemize}
    \item Where is Mattarella [President of the Republic] in all this? [Meaning: what is Mattarella doing to deal with the current situation?] [\textbf{.99}]
    \item  By now we know who Salvini [leader of the right-wing political party] is. The problem is how much is true about this virus [\textbf{.98}]
    \item If we all closed the borders we would all be [considered] fascists [\textbf{.95}]
    \item  I can't get over that someone really believes that the EU will help Italy [\textbf{.79}]
    \item OMG honorable [Member of Parliament] you who called your prime minister a criminal? This is not very credible, right? [.67]
\end{itemize}

\subsection*{Politics Negative Sentiments}

\begin{itemize}
    \item \#Conte [Italy's prime minister] \#wakeup the \#coronavirus advances and reached Lodi [one of the main cities in the red zone] big congratulations to you, perfect checks [\textbf{.93}]
    \item  \#Conte shifts the blame of \#covid contagions to the hospital of Codogno where they have worked hard and they keep working hard with their shifts [\textbf{.51}]
    \item \#coronavirus taught us that the \#nationalhealthservice must be refounded and that the \#TAV [high speed train project that has been subject to heated politically debates] is useless and \#Conte is useless [\textbf{.36}]
    \item But any excuse is valid for this government to give resources to find work for immigrants passing in front of Italians [\textbf{.23}]
    \item Intensive care doctors from several hospitals have called for the silence of politicians [\textbf{.20}]
\end{itemize}

\subsection*{Lockdown Policy Uncertainty}
\begin{itemize}
    \item I am biased but yes we are apparently closed but all of a sudden you will find yourself at the [same] table with us [i.e., in the same situation] [\textbf{.99}]
    \item If they extend the red zone to Lombardy, Italy risks sinking more than it is [already] doing [\textbf{.95}]
    \item It's absurd that the TV information we have in the red zone is the same as the rest of Italy [\textbf{.85}]
    \item Hello Valerio I'm a nurse in the red zone. [Here] the measurements are disproportionate [\textbf{.83}]
    \item the problem is that there is a fear of widening the red zone [\textbf{.76}]
\end{itemize}

\subsection*{Lockdown Policy Negative Sentiments}

\begin{itemize}
    \item I work in psychiatry I can tell you that in the red zone there is madness as they say [\textbf{.97}]
    \item it is yet to be clarified if the match will be played behind closed doors [\textbf{.87}]
    \item but I ask how can you really close everything [?] what a desolation all places [`locali', meaning bars, restaurants...] closed \#coronavirus \#italy \#cremona [\textbf{.60}]
    \item the postponement of Serie A matches scheduled behind closed doors is simply shameful [\textbf{.30}]
    \item it must be that the closed doors mandate in Piedmont [region in North West Italy] is not valid and there are no infected people [\textbf{.19}]
\end{itemize}

\section{Random Choice of the Areas Under Lockdown}\label{app:random}

To further assess the hypothesis of randomized allocation of the lockdown, we test the differences between the red and orange zone with to their observable characteristics, whereas the lack of difference does support its random allocation.
Here, we show that the two groups have comparable pre-treatment demographic and socio-economic characteristics. 

We check for covariate balance among the treated (i.e., red zone) and control (i.e., orange zone) municipalities matching the user location of tweets featured in the sample employed in the DiD estimation (8 red zone and 111 orange zone municipalities). We considered Istat data on a variety of social, economic, and demographic characteristics\footnote{The dataset are collected from \url{https://www.istat.it/it/archivio/241341}, \url{https://www.istat.it/it/archivio/240401} and \url{http://dati-censimentipermanenti.istat.it/.}}. We include the number of residents at the beginning of 2020, the monthly average number of deaths over 2015-2019 for January, February, and March. Moreover, we aggregate census data from 2019 and compute the shares of residents belonging to five occupational categories (Employed, Unemployed, Student, Pension receiver, House worker, and Other), five age cohorts (under 20, 20-39, 40-59, 60-75 and over 75) and six education levels (No education, Primary, Lower Secondary, Upper Secondary, Lower tertiary and Master/Ph.D.). We also collect information on industry and services from 2017, including total output and value added (in euros), number of employees, total staff, and local units. We impute with the group mean missing values on industry and services for two treated cities (Castelgerundo - 1\,473 residents - and Bertonico - 1\,059 residents) and missing entries on monthly average deaths for one control and one treated city (Calendasco - 2\,409 residents - and Bertonico - 1\,059). Before testing, we standardize the variables not converted to proportions, that is, the covariates on mean mortality, industry, services, and total residents.

\begin{figure}
\centering
\includegraphics[width=0.9\textwidth, height=8cm]{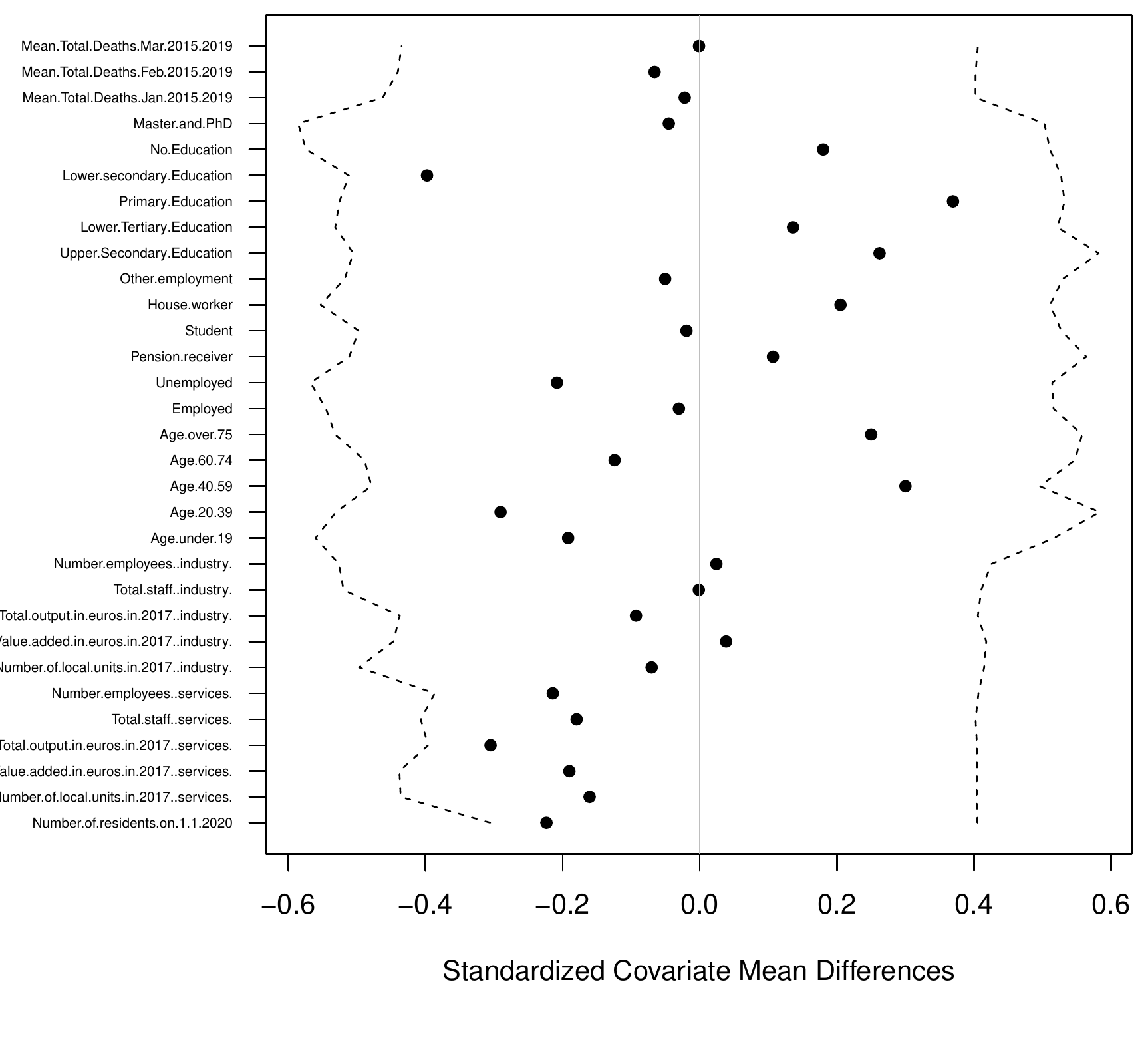}
\caption{Standardized covariate mean difference along with 7.5\% and 92.5\% complete randomization quantiles using 2000 permutations.}
\label{fig:bal_2}
\end{figure}

In Figure \ref{fig:bal_2}, we show for the same covariates the \emph{observed} standardized mean difference, along with the 7.5\% and 92.5\% complete randomization quantiles obtained across 2,000 permutations of the treatment status, assuming complete randomization of the assignment among all units. The quantiles represent the acceptance region of our randomization test with $\alpha  = 0.15$, as we use the standardized covariate mean differences as the test statistic.  For each iteration, the treatment is randomly assigned to units and the standardized mean difference is calculated. An observed standardized difference outside the interval defined by the quantiles is to be considered \emph{extreme} and \emph{unreasonable} under complete randomization of the treatment assignment \autocite{loveplot}. Complete randomization holds for all the considered covariates, as the observed standardized difference is well within the acceptance bounds. 

As the observations we identify as \emph{controls} share the same demographic, geographic, social, and economic composition, as well as the same exposure to the virus, the trends in aggregated and topic-related uncertainty and negative sentiment should not systematically differ between tread and control. Causal identification is achieved by the exogenous shock represented by the policy, such that we exclude selection bias of the treated units, which are balanced and comparable to the control group. Moreover, no other main political event at the city level took place during the period sampled and the area considered does not represent a vital node of the national or regional economy, such that we can exclude any bias coming from unaccounted major events influencing the public's reaction.

\section{Assumptions of the Difference-in-Differences Specification}
\label{sec:assumptions_did}

In our DiD setting, the average treatment effect on the treated in period 1---i.e. ATT(1)---is the causal estimand of interest, defined as
\begin{equation}\label{ATT(1)}
    ATT(1)  = \EX[Y_{ij,1}(1) - Y_{ij,1}(0) | D_{ij} = 1] 
\end{equation}
where  $Y_{ij,t}(0)$ represents the potential outcome of unit $i$ in period $t$ if they remain untreated at time $t$, $Y_{ij,t}(1)$ the potential outcome at time $t$ if they were treated at time $t$. For each unit, we can only observe one of the potential outcomes at each time period.

Because it is possible to observe only $Y_{ij,1}(1)$ for treated units, the identification strategy of the DiD approach imputes the unobserved potential outcomes for the treated units with the \emph{observed} post-treatment outcomes for untreated units that are suitable counterfactuals. The OLS coefficient $\hat{\delta_{1}}$ becomes a consistent estimator of the average treatment effect ATT(1) under the following assumptions \autocite{roth2023whats, RePEc:arx:papers:2105.03737}:


\begin{assumption}
Parallel Trends: 
\begin{equation*}
    \EX[Y_{ij,1}(0) - Y_{ij,0}(0) | D_{ij} = 1] - \EX[Y_{ij,1}(0) - Y_{ij,0}(0) | D_{ij} = 0].
\end{equation*}
\end{assumption}
In other words, without the treatment, the outcome in the treated group would have followed the same trend as the control group. This assumption implies that the units belonging to the two groups, treatment and control, have no inherent systematic differences so that we can rule out any \emph{selection bias} in the estimation of treatment. 

\begin{assumption}
No anticipation of the treatment: 
\begin{equation*}
    Y_{ij,0}(1) = Y_{ij,0}(0) 
\end{equation*}
for all treated units i $D_{ij} = 1$.
\end{assumption}
This assumption implies that before the treatment, the treatment does not impact the outcomes of the treated units---i.e. units cannot modify their outcomes to be selected into the treatment.

\begin{assumption}
\emph{Stable Unit Treatment Values Assumption} (SUTVA)
\begin{equation*}
ATT(1) =  \delta_{1} + \delta^{spill}_{1}(1) - \delta^{spill}_{1}(0) = \delta_{1}
\end{equation*}
\end{assumption}
where $\delta^{spill}_{1}(1)$ is the average spillover effect on treated units and $\delta^{spill}_{1}(0)$ the average spillover non-treated units in period 1, which are set to zero under SUTVA. The treatment assignment or the potential outcome of a unit cannot affect the potential outcome of another unit.

Under these assumptions and random sampling, it follows that:
\begin{equation}
    ATT(1)  = \EX[Y_{ij,1} - Y_{ij,0} | D_{ij} = 1] - \EX[Y_{ij,1} - Y_{ij,0} | D_{ij} = 0]
\end{equation}

\section{Testing DiD Assumptions}\label{test_assumptions}

\subsection{Pre-existing trends}
\label{sec: pre-trends}
The validity of a DID model rests on the key identification assumption of \emph{parallel trends}.
We test whether the assumption holds by regressing the DID model on \emph{m} leads and \emph{q} lags of the treatment variable over multiple periods \autocite{pischke}: 
\begin{equation}\label{pt_eq}
Y_{ij,t} = \alpha + \gamma_{j} + \sum^{q}_{l=-m}\lambda_{l}\bbone[T_{ij} = l] + \sum^{q}_{l=-m}\delta_{l}(\bbone[T_{ij} = l] \times D_{ij}) + \epsilon_{ij,t} 
\end{equation}

If treatment and control show the same outcome trend in the pre-treatment period, the $\delta_{l}$ coefficient should be small in magnitude and non-significant for $l < 0$, that is, for the $l$th period occurring before the lockdown. 
We take as baseline the \emph{pre-treatment} period just before the start of the treatment, between February 1, and February 19, 2020. We define time dummies indicating January 2020, the post-treatment period (February 23, 2020 to March 6, 2020), and the period March 9-March 21, 2020. Then, $\delta_{-1}$ is defined as the coefficient on the interaction of the treatment indicator with the January dummy, $\delta_{0}$ with the post-treatment period dummy, and $\delta_{1}$ with the post-post-treatment dummy.

\begin{figure}[H]
  \begin{subfigure}[h]{0.4\textwidth}
    \includegraphics[width=0.8\textwidth]{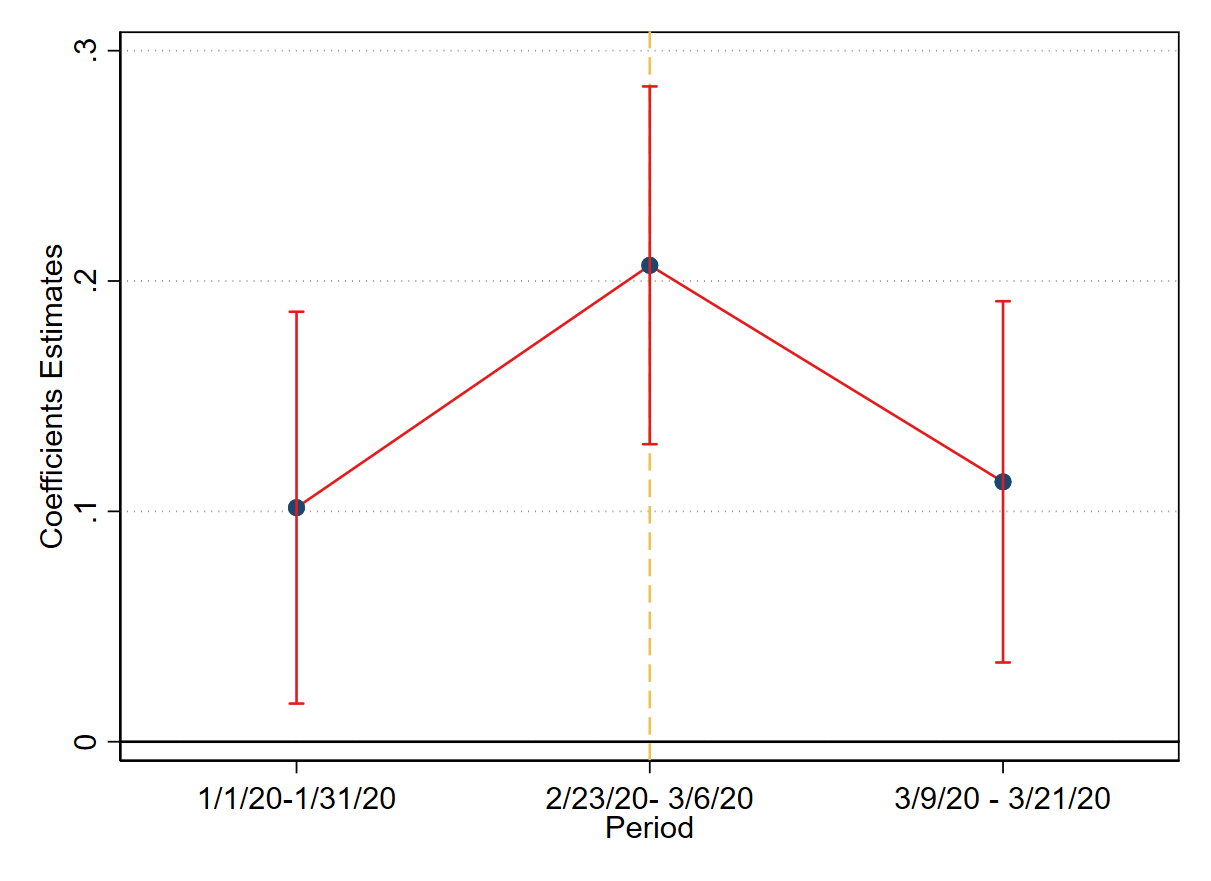}
    \caption{Uncertainty}
    \label{fig:pt_u1}
  \end{subfigure}
  \hfill
 \begin{subfigure}[h]{0.4\textwidth}
    \includegraphics[width=0.8\textwidth]{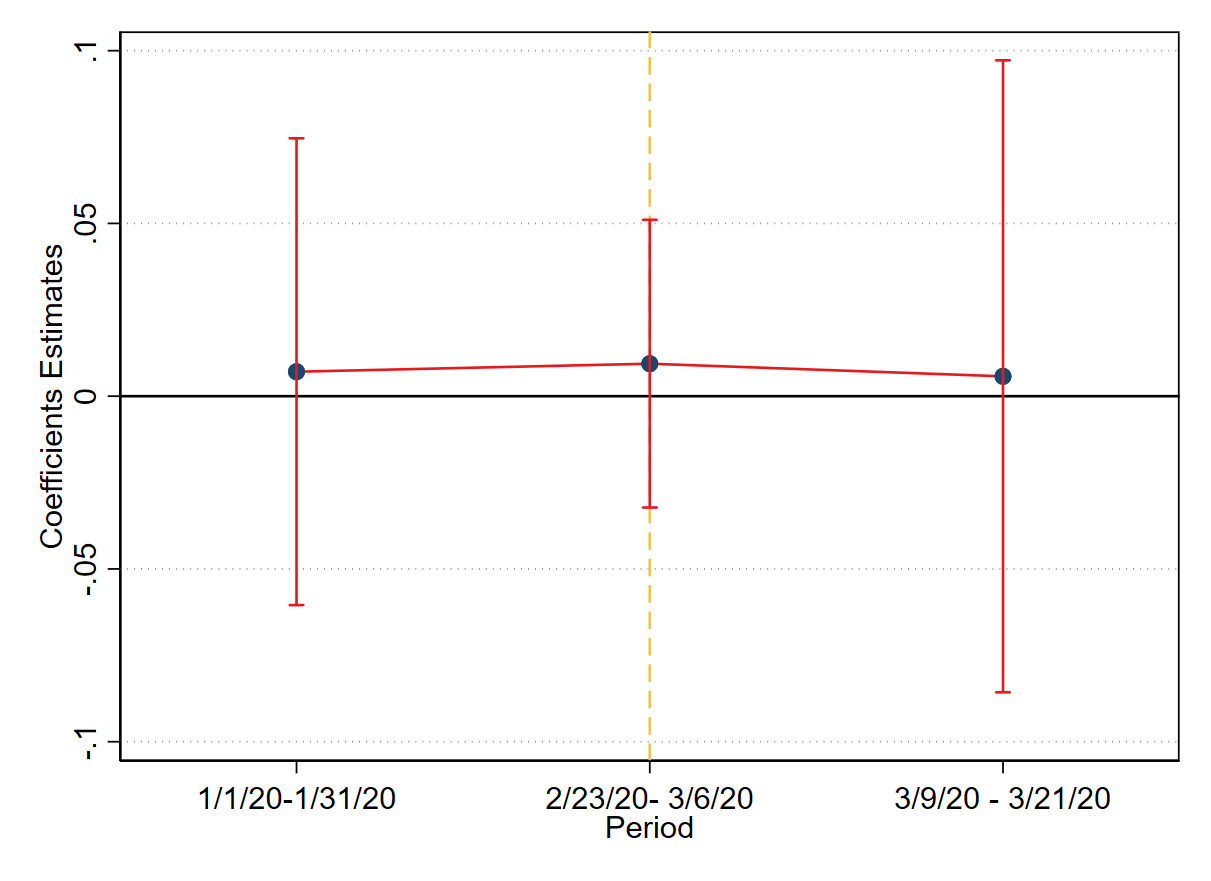}
    \caption{Negative Sentiment}
    \label{fig:pt_u2}
  \end{subfigure}
  \begin{subfigure}[h]{0.4\textwidth}
    \includegraphics[width=0.8\textwidth]{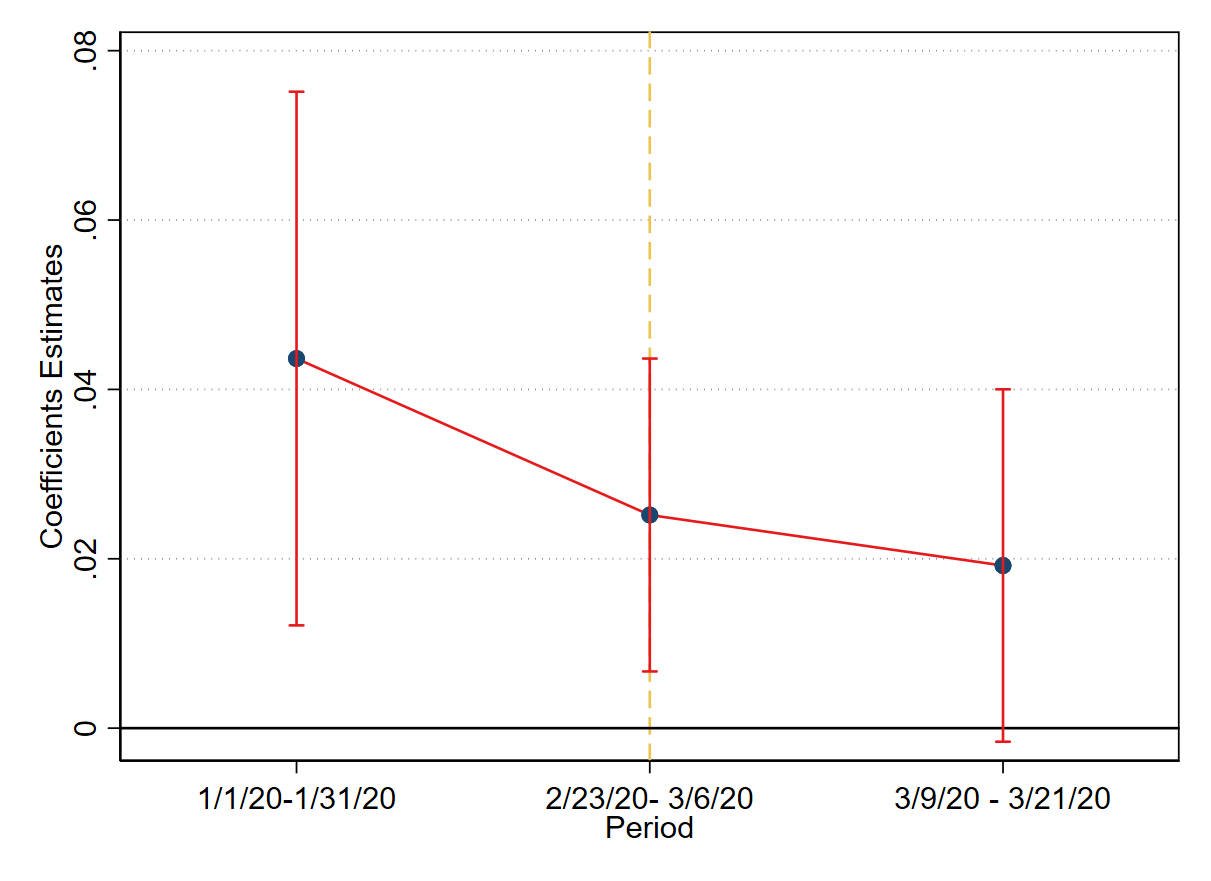}
    \caption{Uncertainty-Economics}
    \label{fig:pt_u3}
  \end{subfigure}
  \hfill
 \begin{subfigure}[h]{0.4\textwidth}
    \includegraphics[width=0.8\textwidth]{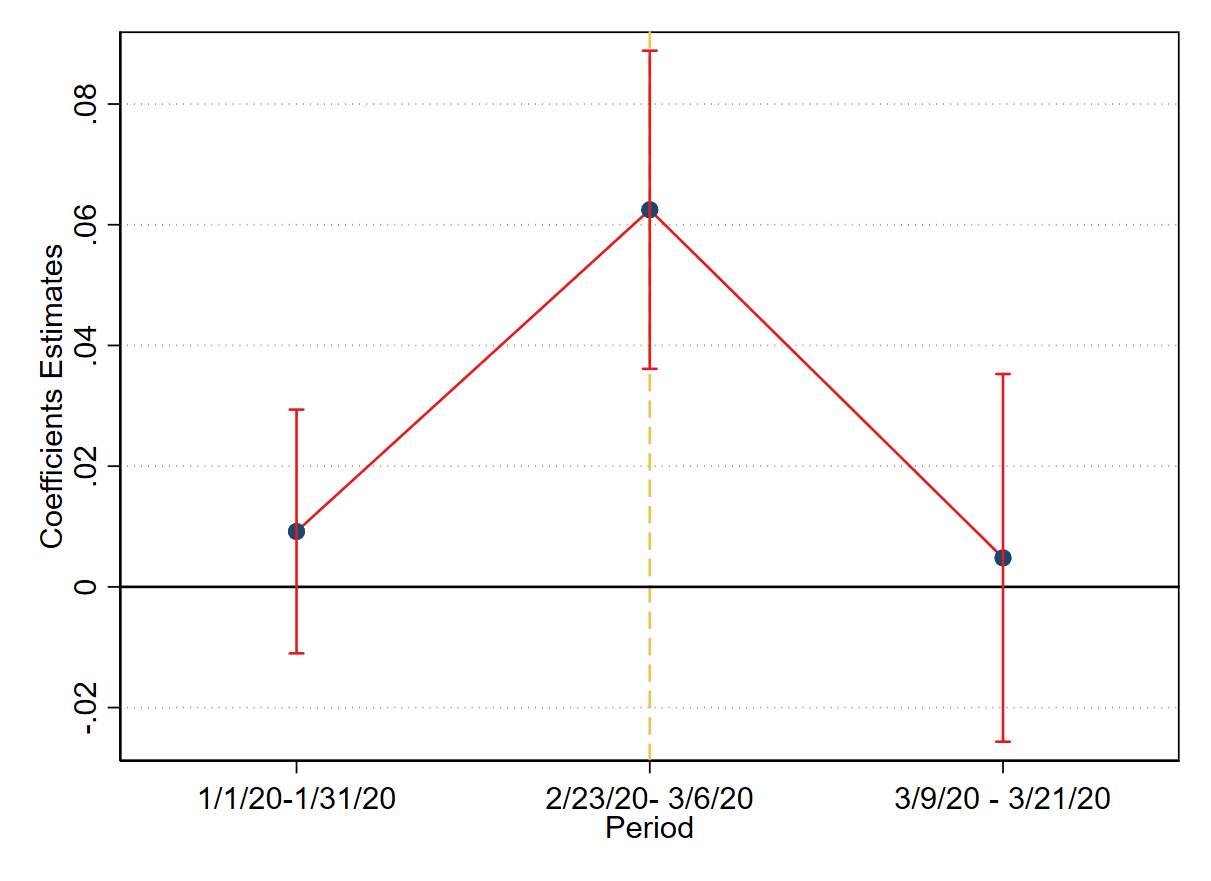}
    \caption{Uncertainty-Health}
    \label{fig:pt_u4}
  \end{subfigure}
  \begin{subfigure}[h]{0.4\textwidth}
    \includegraphics[width=0.8\textwidth]{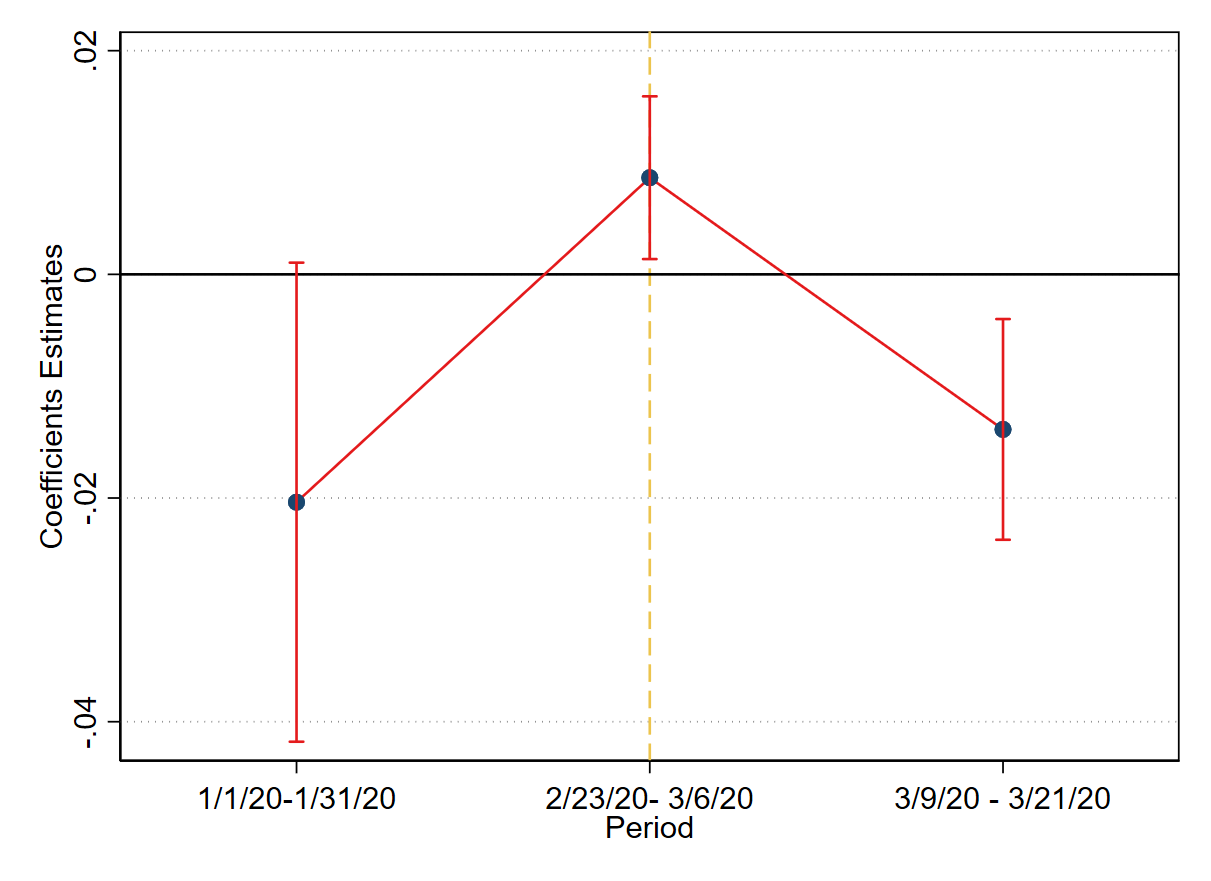}
    \caption{Uncertainty-Politics}
    \label{fig:pt_u5}
  \end{subfigure}
  \hfill
 \begin{subfigure}[h]{0.4\textwidth}
    \includegraphics[width=0.8\textwidth]{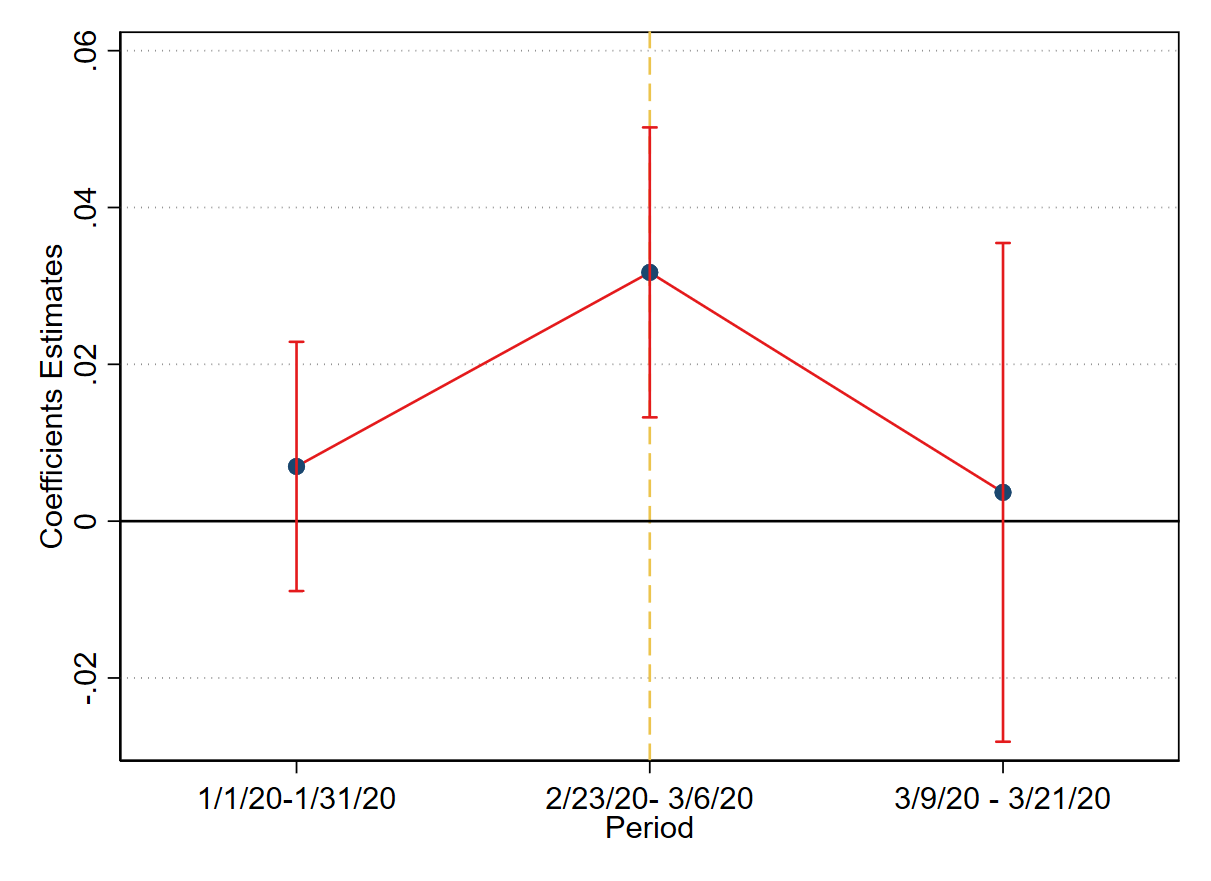}
    \caption{Uncertainty-Policy}
    \label{fig:pt_u6}
  \end{subfigure}
    \begin{subfigure}[h]{0.4\textwidth}
    \includegraphics[width=0.8\textwidth]{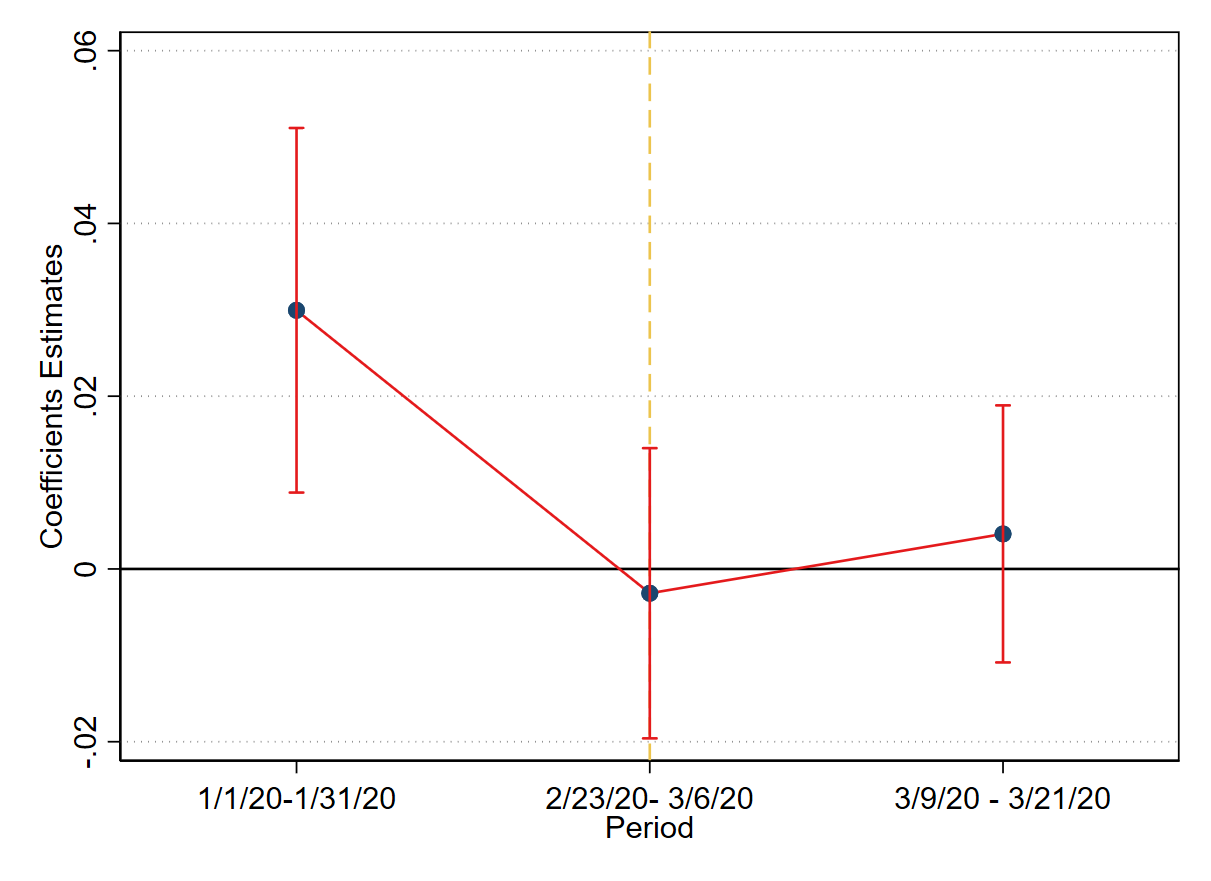}
    \caption{Negative Sentiment-Economics}
    \label{fig:pt_u7}
  \end{subfigure}
  \hfill
 \begin{subfigure}[h]{0.4\textwidth}
    \includegraphics[width=0.8\textwidth]{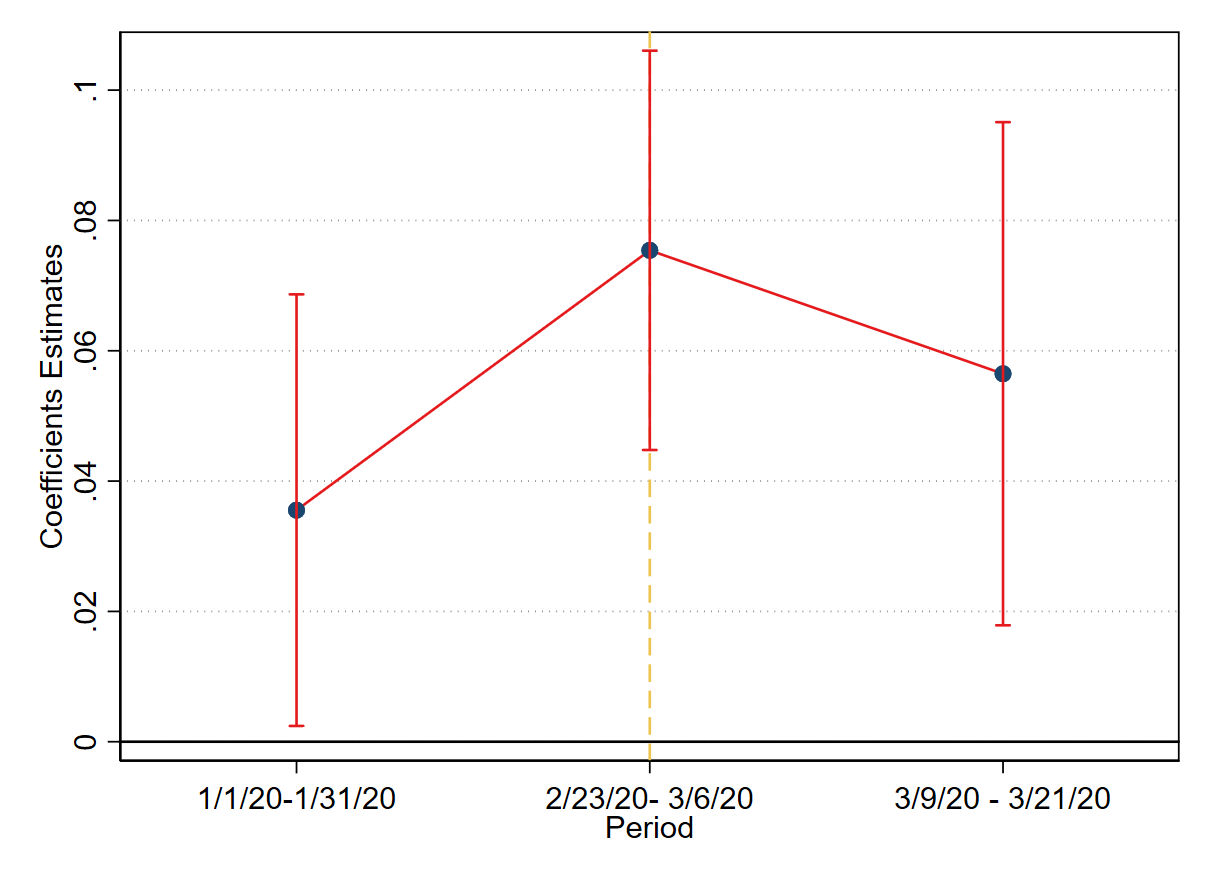}
    \caption{Negative Sentiment-Health}
    \label{fig:pt_u8}
  \end{subfigure}
  \begin{subfigure}[h]{0.4\textwidth}
    \includegraphics[width=0.8\textwidth]{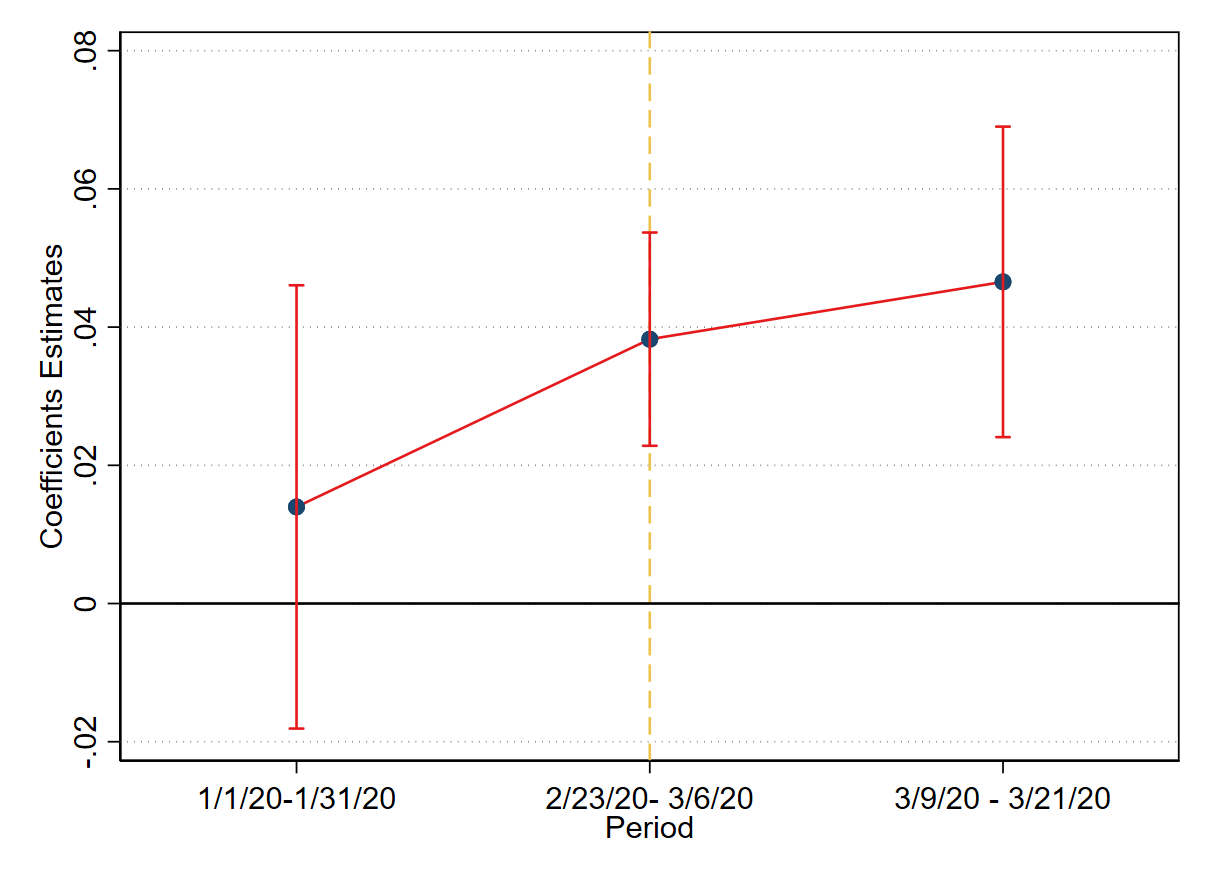}
    \caption{Negative Sentiment-Politics}
    \label{fig:pt_u9}
  \end{subfigure}
  \hfill
 \begin{subfigure}[h]{0.4\textwidth}
    \includegraphics[width=0.8\textwidth]{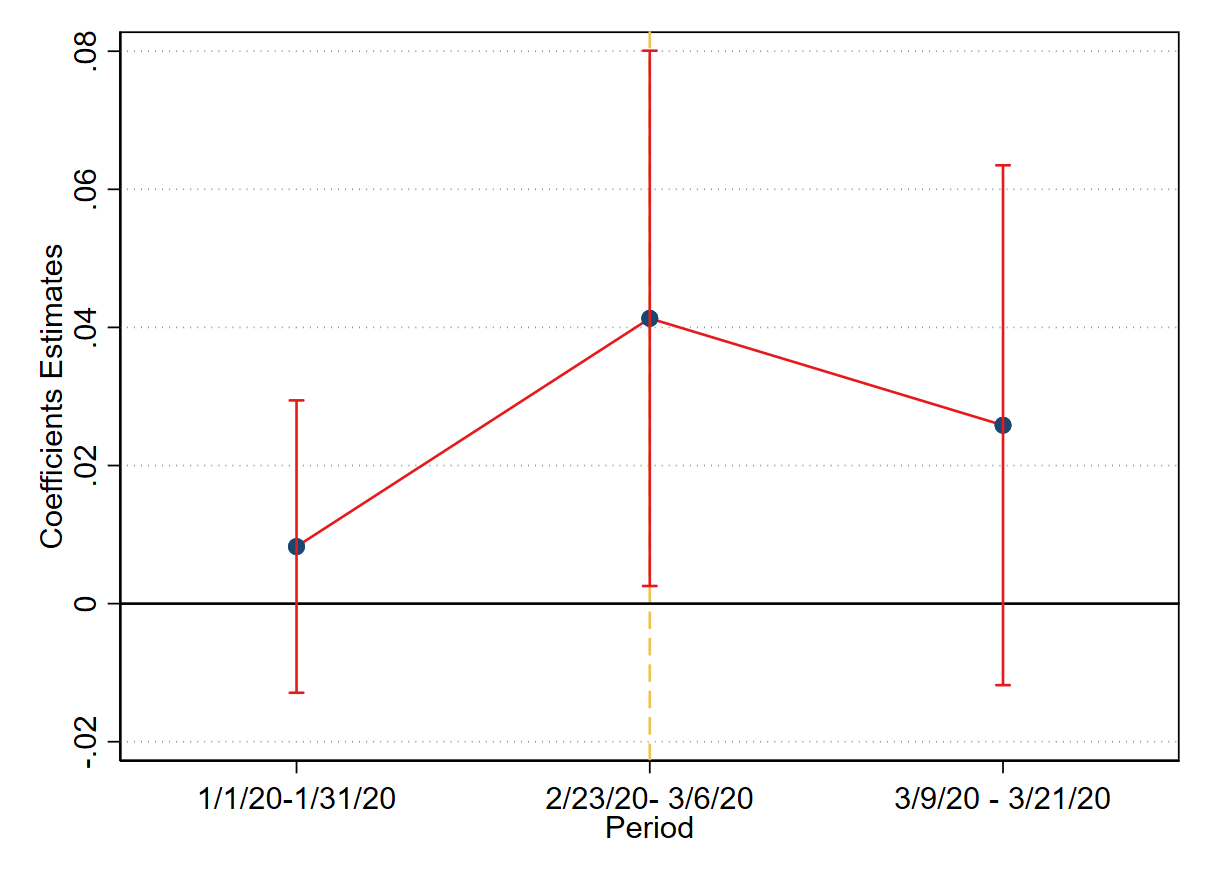}
    \caption{Negative Sentiment-Policy}
    \label{fig:pt_u10}
  \end{subfigure}
  \caption{Coefficient Estimates and Confidence Intervals (\% 95) of interaction between the treatment variable and time dummies. The dependent variable is Share of tweets classified as \emph{Uncertainty} and \emph{Negative Sentiment} aggregated and grouped by Topic. The baseline period starts from the 1rst of February and ends on the 19th of February.}
  \label{fig:pt_plots}
\end{figure}

In Figure \ref{fig:pt_plots}, we see that the magnitude and significance of coefficients are in line with previous findings for uncertainty towards health, uncertainty and negative sentiment related to politics and the policy, as the difference in pre-treatment trends is not statistically significant. On the other hand, we reject the hypothesis of parallel pre-treatment trends for aggregated uncertainty and negative sentiments towards to health.


Although it is common practice to perform statistical hypothesis testing of the differences in pre-treatment trends between treated and control groups, the assumption still fundamentally relies on unobservable quantities, that is, the outcome of the treatment group in absence of the treatment. Not rejecting the parallel trends hypothesis is neither a necessary nor a sufficient condition of the DID model, which still remains untestable and based on unobservable counterfactual quantities. The researcher should always provide a robust, logical explanation of why the parallel trend assumption should hold, as the conditions for parallel counterfactual trends hold only by the logical reasoning underpinning the empirical design \autocite{kahn-lang}.

\subsection{Sensitivity analysis with bounds on pre-trends}\label{honestdid_sect}


Next, we assess the robustness of the estimates of $\delta_{0}$ in equation \ref{pt_eq} -- i.e. the effect of the first lockdown on those living in the \emph{red zone}, as we allow for some violation of the parallel trends conditions following \textcite{RambachanRoth23}. 
With this approach, the size of post-treatment violation of parallel trends is bounded, as we assume that it cannot be larger than $\bar{M}$ times the largest pre-treatment violation \autocite{roth2023whats}.

We concentrate on the effect of the lockdown on those first treated by the measures, therefore we drop all observations from after the extension of the measure at the national level to avoid dealing within a staggered setting - as all units become treated on March 9 and so the control group is left empty. We perform sensitiviy analysis on $\delta_{0}$ using the HonestDiD package\footnote{\url{https://github.com/asheshrambachan/HonestDiD}}, that is, we allow for different values of $\bar{M}$ and report the confidence intervals of $\delta_{0}$ for changing the value of $\bar{M}$. As the effect becomes insignificant for some values of $\bar{M}$, then we obtain a significant result that is robust to violation of parallel trends that can as large as $\bar{M}$ times the pre-treatment violation in trends.

\begin{figure}[H]
  \begin{subfigure}[h]{0.3\textwidth}
    \includegraphics[width=\textwidth]{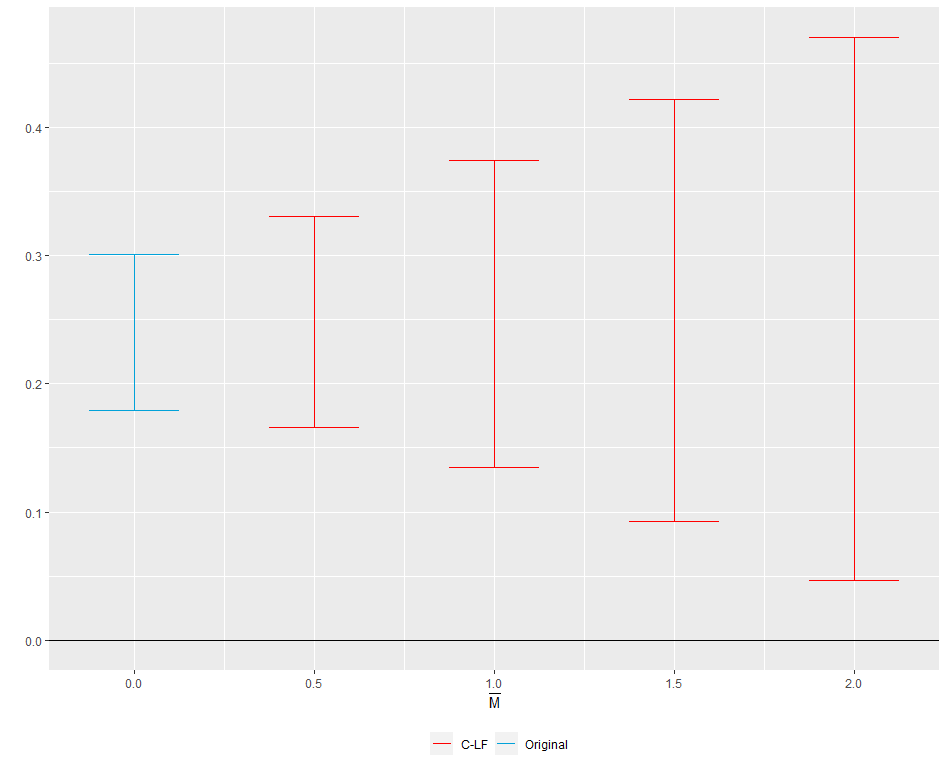}
    \caption{Uncertainty}
    \label{fig:user_FEs_honestdid_1aa}
  \end{subfigure}
  \hfill
  \begin{subfigure}[h]{0.3\textwidth}
    \includegraphics[width=\textwidth]{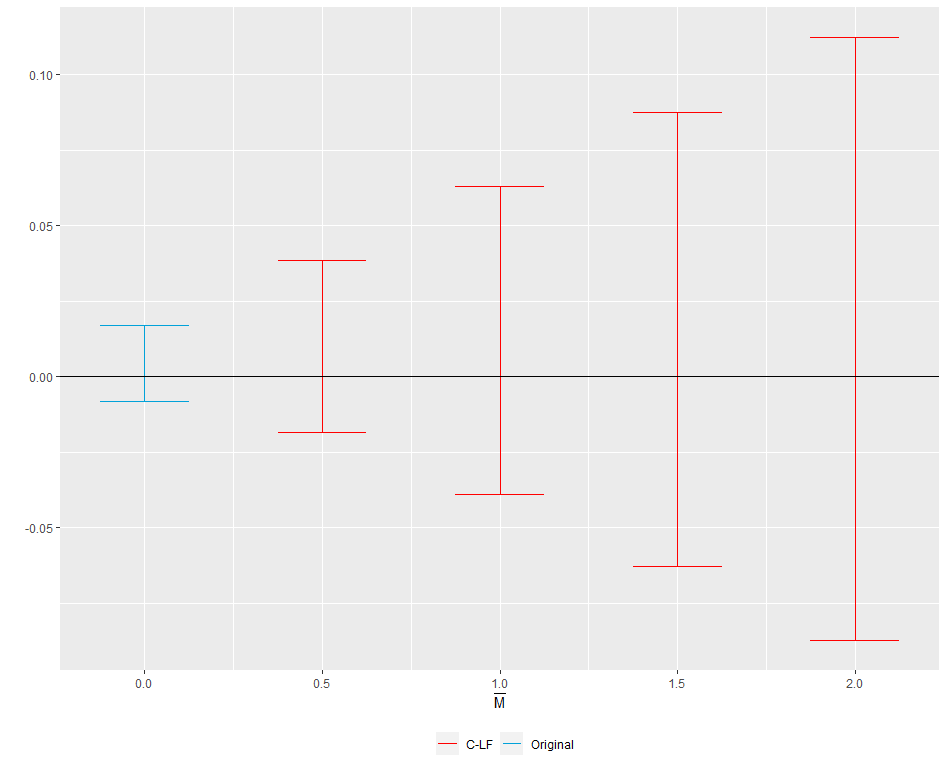}
    \caption{Uncertainty-Economics}
    \label{fig:user_FEs_honestdid_2a}
  \end{subfigure}
  \hfill
 \begin{subfigure}[h]{0.3\textwidth}
    \includegraphics[width=\textwidth]{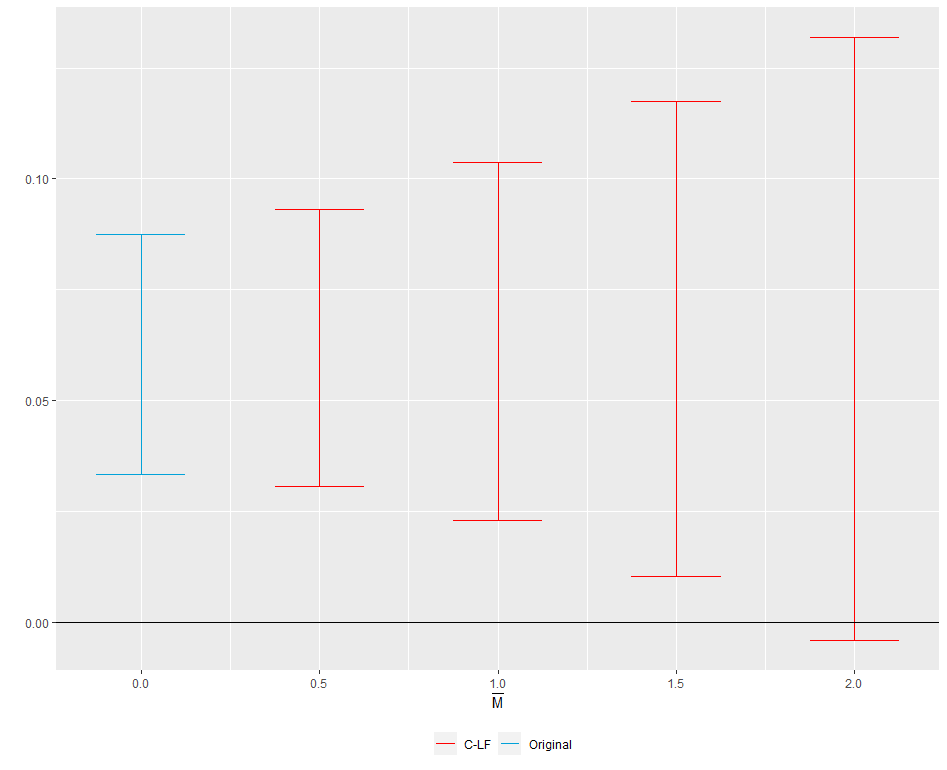}
    \caption{Uncertainty-Health}
    \label{fig:user_FEs_honestdid_2b}
  \end{subfigure}
  \begin{subfigure}[h]{0.3\textwidth}
    \includegraphics[width=\textwidth]{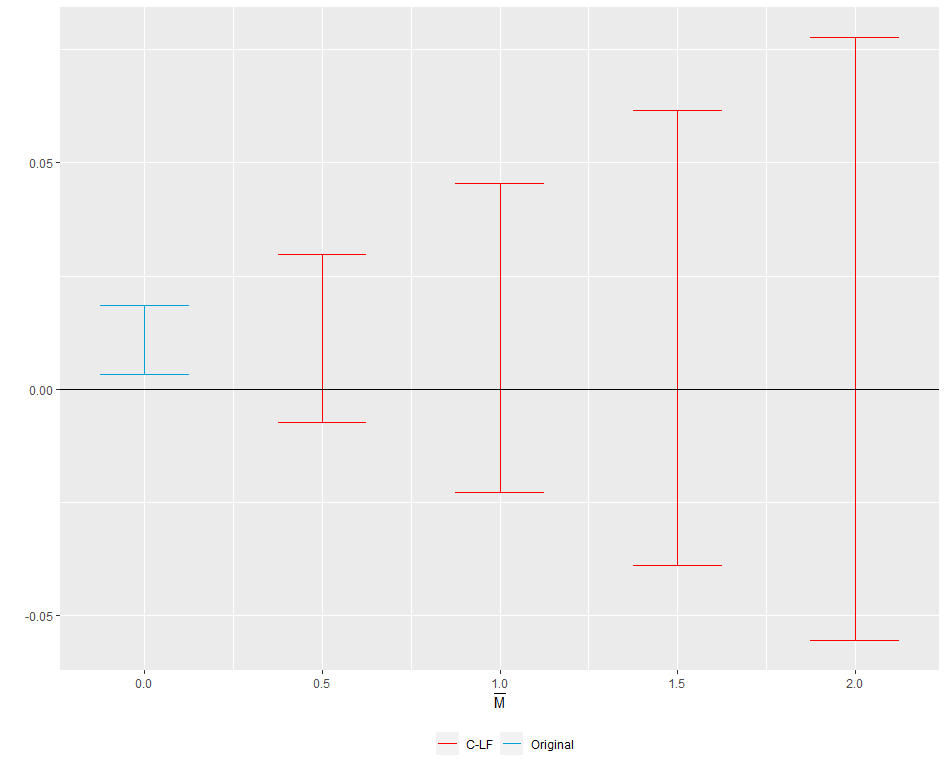}
    \caption{Uncertainty-Politics}
    \label{fig:user_FEs_honestdid_2c}
  \end{subfigure}
  \hfill
 \begin{subfigure}[h]{0.3\textwidth}
    \includegraphics[width=\textwidth]{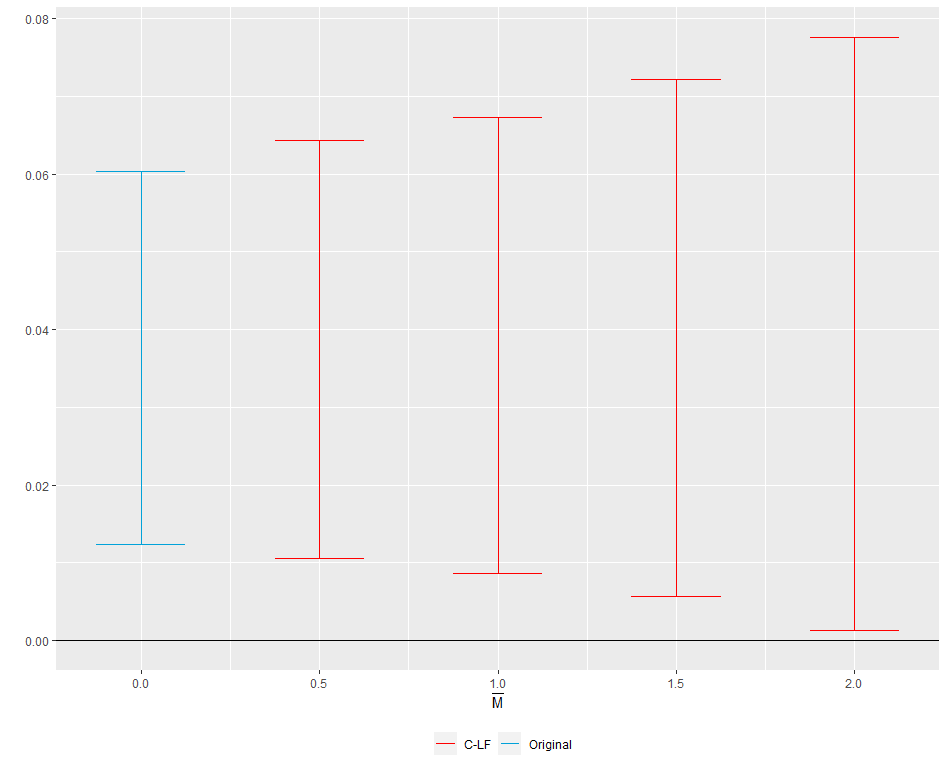}
    \caption{Uncertainty-Policy}
    \label{fig:user_FEs_honestdid_2d}
  \end{subfigure}
  \caption{Sensitivity analysis using relative magnitudes restrictions of $\delta_{0}$ of equation \ref{pt_eq} for \emph{Uncertainty} aggregated and grouped by Topic. The baseline period starts from the 1rst of February and ends on the 19th of February.}
  \label{fig:honestdid_2}
\end{figure}

\begin{figure}[H]
 \begin{subfigure}[h]{0.3\textwidth}
    \includegraphics[width=\textwidth]{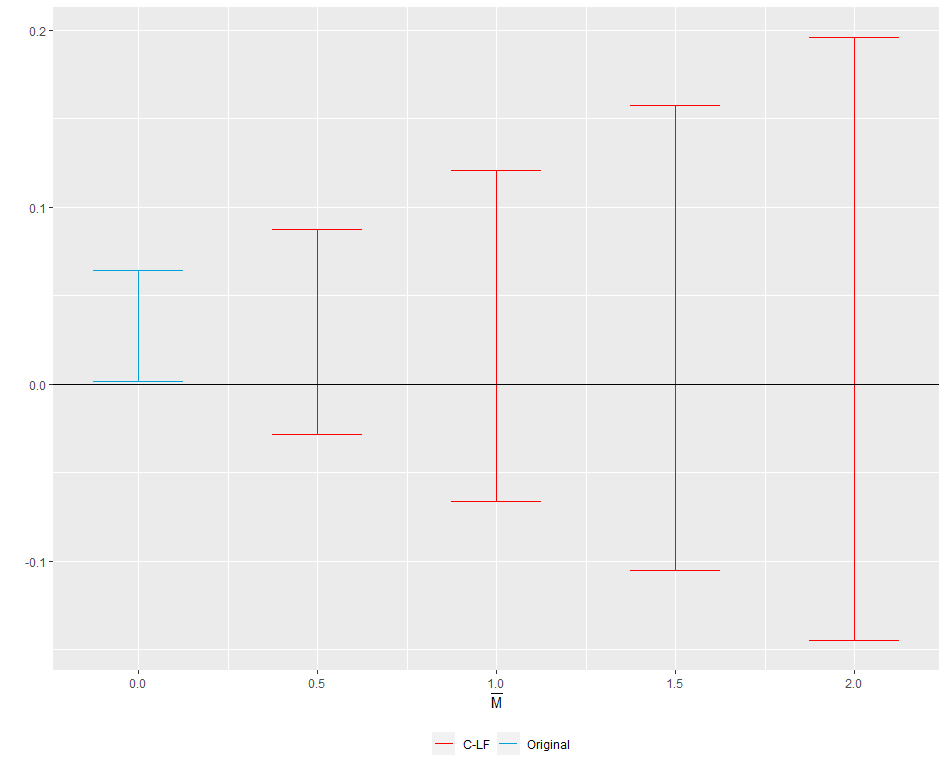}
    \caption{Negative Sentiment}
    \label{fig:user_FEs_honestdid_1b}
  \end{subfigure}
  \hfill
    \begin{subfigure}[h]{0.3\textwidth}
    \includegraphics[width=\textwidth]{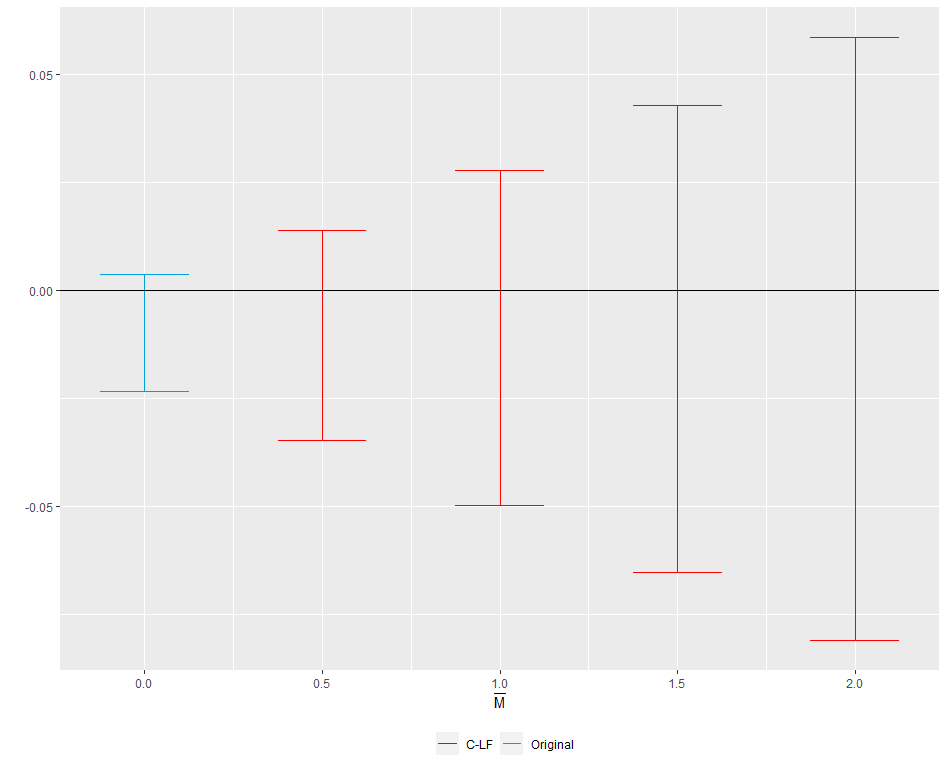}
    \caption{Negative Sentiment-Economics}
    \label{fig:user_FEs_honestdid_3a}
  \end{subfigure}
  \hfill
 \begin{subfigure}[h]{0.3\textwidth}
    \includegraphics[width=\textwidth]{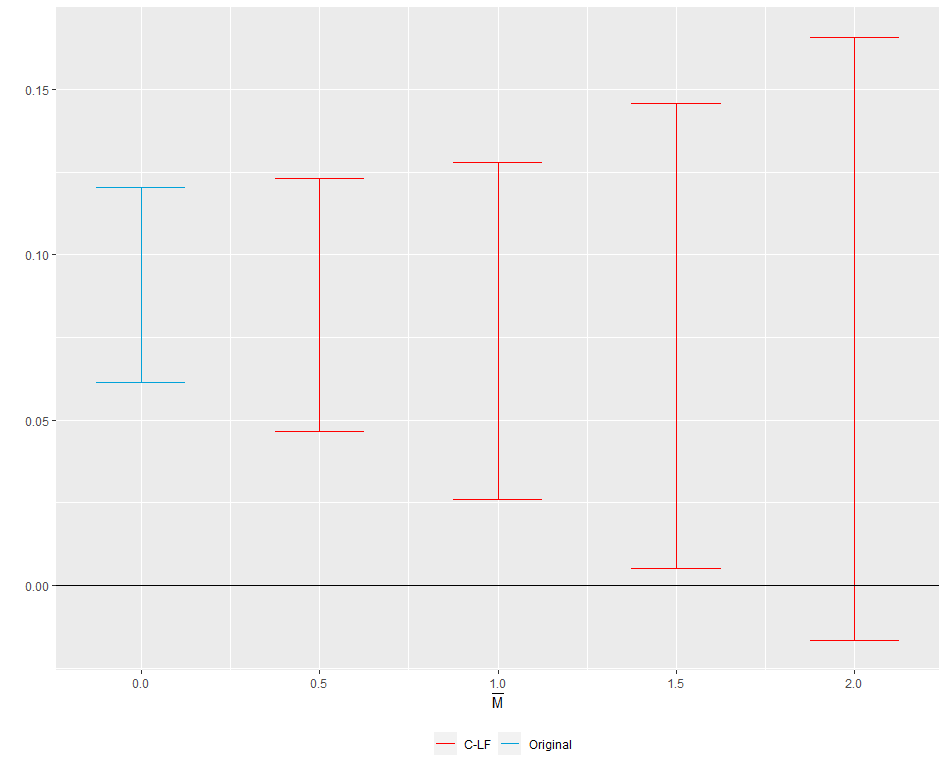}
    \caption{Negative Sentiment-Health}
    \label{fig:user_FEs_honestdid_3b}
  \end{subfigure}
  \begin{subfigure}[h]{0.3\textwidth}
    \includegraphics[width=\textwidth]{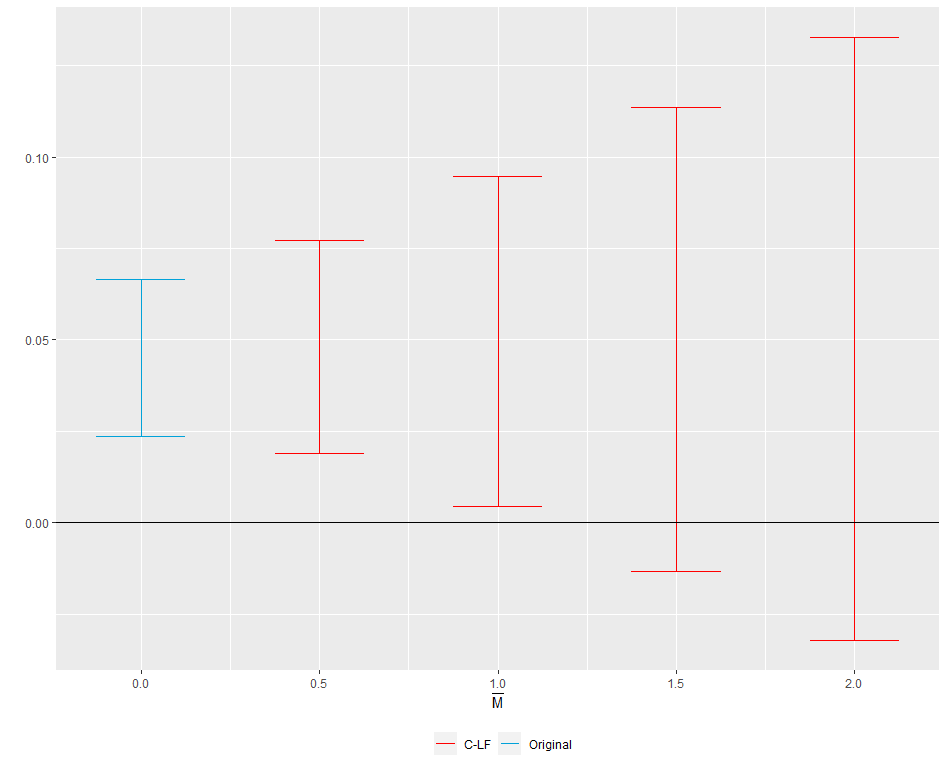}
    \caption{Negative Sentiment-Politics}
    \label{fig:user_FEs_honestdid_3c}
  \end{subfigure}
  \hfill
 \begin{subfigure}[h]{0.3\textwidth}
    \includegraphics[width=\textwidth]{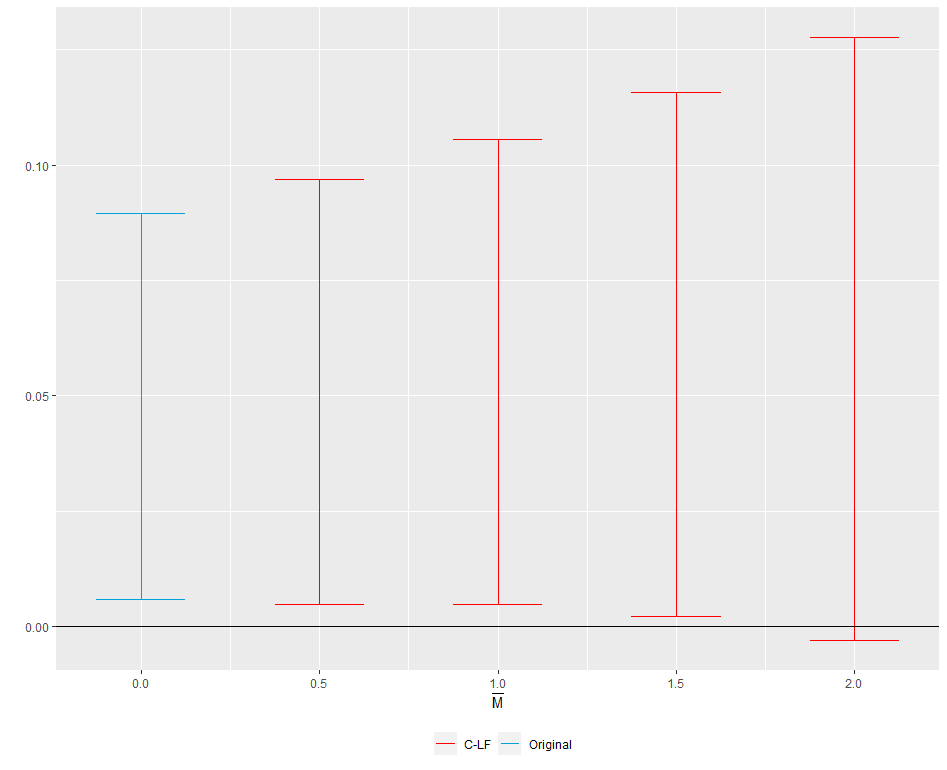}
    \caption{Negative Sentiment-Policy}
    \label{fig:user_FEs_honestdid_3d}
  \end{subfigure}
  \caption{Sensitivity analysis using relative magnitudes restrictions of $\delta_{0}$ of equation \ref{pt_eq} for \emph{Negative Sentiment} aggregated and grouped by Topic. The baseline period starts from the 1rst of February and ends on the 19th of February.}
  \label{fig:honestdid_3}
\end{figure}

The results of all outcomes of interest are reported in Figure \ref{fig:honestdid_2} and Figure \ref{fig:honestdid_3}. The test returns 95\% \emph{robust} confidence intervals for different values of $\bar{M}$\footnote{We refer to the original paper by \textcite{RambachanRoth23} for a discussion of the method.}. 

By allowing up to 1.5 times the maximal violation in pre-treatment trends we still retain a robust significant effect for uncertainty and negatives sentiments towards health and the policy, and up to 1 time for negative sentiments towards politics.



On the other hand, the effect on uncertainty related to politics is not robust to any violations of parallel trends, as depicted in Figure \ref{fig:user_FEs_honestdid_2c}. Therefore, the effect is only significant as we assume zero violations of parallel trends, in line with the usual parallel trend assumption of the DiD model.

\section{Multiple Test Adjusted P-values}\label{adj_pval_sect}

We report the Benjamini-Hochberg adjusted pvalues \autocite{BH} for multiple hypothesis testing for false discovery rate (with uncorrected p-values in parenthesis). We jointly test the p-values of the model for Aggregate uncertainty (sentiment) with additional regressions for two new binary variables representing the correlated categories of \emph{certainty} and \emph{medium} uncertainty (\emph{positive} and \emph{medium} sentiment), as there are likely correlations amongst the p-values of the three regression models that may have been inflating our significance. Considering uncertainty and negative sentiment grouped by possible topics, the p-values of the four topic-related independent regressions are adjusted together. 

In Table \ref{adj_p2}, we confirm all main findings of the baseline estimation as the treatment effect (\emph{red} $zone$ x $post = 1$) retains its statistical significance on uncertainty towards economics, health and politics and negative sentiment on health and politics. This indicates that the original p-value of the policy effect on public reaction was not inflated by correlations among the regression models, leading to an incorrect inference. 




\begin{table}[H]\centering
\def\sym#1{\ifmmode^{#1}\else\(^{#1}\)\fi}
\caption{Benjamini Hochberg (1995) adjusted p-values (original p-values in parenthesis)\label{adj_p2}}
\begin{adjustbox}{max width=\textwidth}
\begin{tabular}{lrrrrrrrrrr}
\toprule 
\multicolumn{1}{c}{}&\multicolumn{5}{c}{\textbf{Uncertainty}}&\multicolumn{5}{c}{\textbf{Negative Sentiment}}\\
\toprule
         &\multicolumn{1}{c}{Aggregate}&\multicolumn{1}{c}{Economics}&\multicolumn{1}{c}{Health}&\multicolumn{1}{c}{Politics}&\multicolumn{1}{c}{Policy}&\multicolumn{1}{c}{Aggregate}&\multicolumn{1}{c}{Economics}&\multicolumn{1}{c}{Health}&\multicolumn{1}{c}{Politics}&\multicolumn{1}{c}{Policy}\\
\midrule
post = 1     & 0.01 & 0.02 & 0.00  & 0.11  &  0.00 & 0.10 & 0.95 & 0.00 & 0.11 & 0.00\\
           & (0.00) & (0.01) & (0.00)& (0.07) & (0.00) & (0.08) & (0.95) & (0.00)& (0.07) & (0.01)\\
post = 2   & 0.01 & 0.00& 0.00 & 0.72 & 0.00 & 0.00 & 0.28 & 0.00  & 0.02 & 0.00\\
            & (0.00) & (0.00) & (0.00)& (0.54) & (0.00) & (0.03) & (0.26) & (0.00)& (0.01) & (0.00)\\
\textbf{red zone = 1 x post = 1}  & 0.03 & 0.84 & 0.00 & 0.00 & 0.01 & 0.63 & 0.17 & 0.00 & 0.00 & 0.02 \\
& (0.02) & (0.78) & (0.00)& (0.00) & (0.00) & (0.63) & (0.14) & (0.00)& (0.00) & (0.01)\\
red zone = 1 x post = 2 & 0.57 & 0.72 & 0.94 & 0.41 & 0.99 & 0.63 & 0.19 & 0.10  & 0.00 & 0.12 \\
            & (0.49) & (0.59) & (0.88)& (0.10) & (0.29) & (0.99) & (0.16) & (0.07)& (0.00) & (0.09)\\
\bottomrule
\end{tabular}
\end{adjustbox}
\end{table}

\section{Placebo Test}\label{placebo_sect}

We perform a \emph{placebo} test to see if we can retrieve the effects of the lockdown when the implementation of quarantine measures is not unexpected by the public.

We argue that in those cities where the perceived risk of the danger of the virus is \emph{similar} to the first 10 cities under quarantine,  after the national lockdown there should not be any particular different reaction compared to cities with a totally dissimilar perceived risk, since the announcement of the nation-wide measures were easily anticipated and expected. We take the percentage increase of total deaths at the monthly and city level available on the Istat archives\footnote{Istat data available at https://www.istat.it/it/archivio/240401, under ``4 maggio 2020''} as a proxy variable for the perceived risk related to the pandemics. For each city, the percentage increase over a given month in 2020, is computed as the total number of deaths in that month minus the mean of the total number of deaths in the same month over the years 2015-2019 divided by the mean of the total deaths of that month over the period 2015-2019. For instance, if we take the month of January 2020, we compute the total number of deaths in January 2020 minus the mean of the total number of deaths of January over the years 2015-2019, divided the mean of the total deaths of January over 2015-2019.

We select tweets geo-located in municipalities from the North of Italy only (from a larger dataset of tweets at the national level), and exclude those municipalities used as controls for the Diff-In-Diff models.

\begin{figure}[H]
\centering
\includegraphics[scale = 0.3]{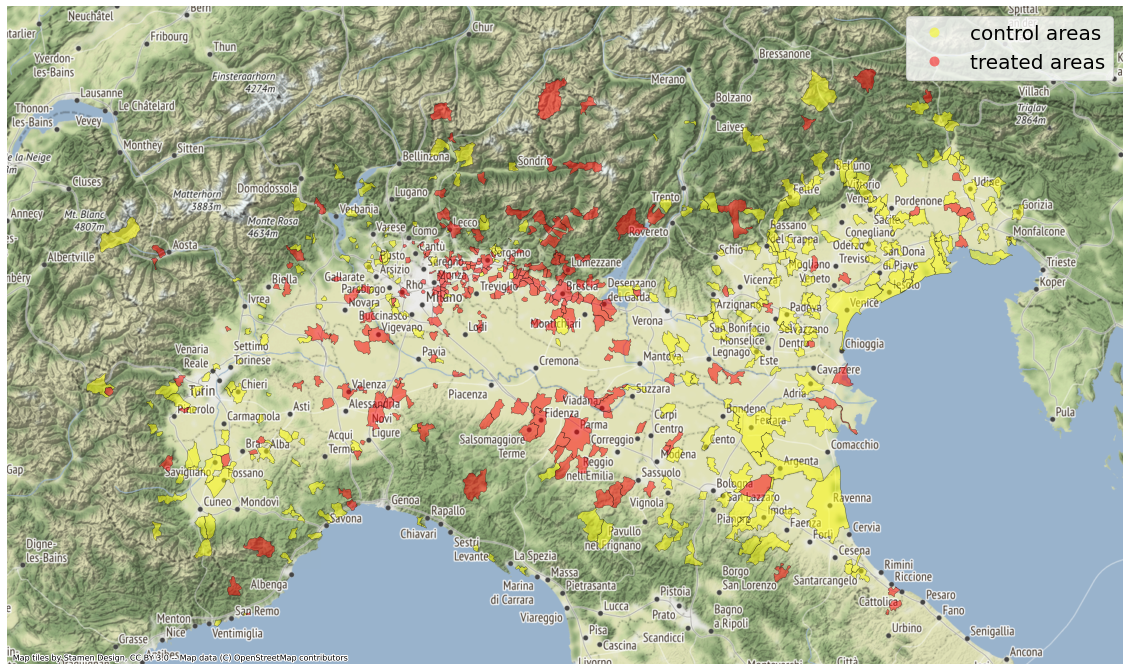}
\caption{Placebo test cities under lockdown (the treated group - in \emph{red}) and controls (in \emph{yellow}).}
\label{fig:pl_1}
\end{figure}

For each city, we consider the percentage increase of deaths of January and February 2020. Each location is therefore represented by a two-dimensional vector and we wish to compare this vector to another two-dimensional array of the monthly percentage increases averaged over the ten cities undergoing the first lockdown. The overall distance between a city and the averages over the ten under the first lockdown is computed based on the Euclidean distance. This allows us to define the binary variable \emph{sim} indicating whether the perception of risk in the city is close to that of the ten under quarantine: the positive label 1 indicates all those cities with Euclidean distance less or equal to the 10th percentile of the distribution of Euclidean distance values in the tweets sample (i.e. less or equal to 316.8821), the negative label 0 stands for all cities with distance equal or above the 90th percentile (i.e. above 424.1583). The cities and therefore observations in the middle are removed to obtain a clean effect and not end up comparing two groups with gradually homogeneous increase in total deaths (Figure \ref{fig:pl_1}). The cities with positive (negative) similarity are 231 (295) and account for 6\,967 (7\,082) observations.

We compare the national lockdown effect on public interests displayed on Twitter between these set of cities with $sim = 1$  and the municipalities featuring $sim = 0$ coming from the same regions. We include a time dummy $Post$ that equals 1 starting from March 9, following the government's decision to extend the measures nation-wide. Again, we drop all observations from March 8, as some municipalities enter quarantine one day before the national lockdown. We also remove March 7, the day of the announcement of the one-day lockdown for 14 provinces in Northern Italy. To control for unobserved time-invariant heterogeneity arising from the large number of different and far-apart municipalities included in the sample, we include municipality level fixed-effects.

From Table \ref{pl_2}, we have no significant effect of the treatment, suggesting that indeed the significant estimates we found in the baseline models are robust to any anticipatory effect of the treatment.


\begin{table}[H]\centering
\def\sym#1{\ifmmode^{#1}\else\(^{#1}\)\fi}
\caption{Placebo Test regression table for \emph{Uncertainty} and \emph{Negative Sentiment}, aggregated and grouped by topics, with municipality level fixed effects (omitted).\label{pl_2}}
\begin{adjustbox}{max width=\textwidth}
\begin{tabular}{lrrrrrrrrrr}
\toprule
&\multicolumn{5}{c}{Uncertainty}&\multicolumn{5}{c}{Negative Sentiment}\\
\toprule
                    &\multicolumn{1}{c}{(1)}&\multicolumn{1}{c}{(2)}&\multicolumn{1}{c}{(3)}&\multicolumn{1}{c}{(4)}&\multicolumn{1}{c}{(5)}&\multicolumn{1}{c}{(6)}&\multicolumn{1}{c}{(7)}&\multicolumn{1}{c}{(8)}&\multicolumn{1}{c}{(9)}&\multicolumn{1}{c}{(10)}\\
                    &\multicolumn{1}{c}{Aggregate}&\multicolumn{1}{c}{Economics}&\multicolumn{1}{c}{Health}&\multicolumn{1}{c}{Politics}&\multicolumn{1}{c}{Policy}&\multicolumn{1}{c}{Aggregate}&\multicolumn{1}{c}{Economics}&\multicolumn{1}{c}{Health}&\multicolumn{1}{c}{Politics}&\multicolumn{1}{c}{Policy}\\
\midrule
\addlinespace
post=1              &     -0.0960\sym{***}&    -0.00796         &     -0.0771\sym{***}&     -0.0134\sym{*}  &     -0.0248\sym{**} &     -0.0434\sym{**} &    -0.00159         &     -0.0397\sym{**} &     -0.0127         &     0.00225         \\
                    &     (-6.73)         &     (-1.12)         &     (-5.38)         &     (-2.15)         &     (-3.20)         &     (-2.89)         &     (-0.29)         &     (-3.20)         &     (-1.70)         &      (0.49)         \\
\addlinespace
sim=1 $\times$ post=1&      0.0475         &      0.0105         &      0.0476         &     0.00491         &      0.0122         &     -0.0322         &     -0.0102         &     -0.0145         &    -0.00556         &   -0.000633         \\
                    &      (1.76)         &      (0.93)         &      (1.89)         &      (0.53)         &      (1.17)         &     (-1.61)         &     (-1.24)         &     (-0.77)         &     (-0.62)         &     (-0.10)         \\
\addlinespace
Constant            &       0.564\sym{***}&     0.00530         &       0.385\sym{***}&     0.00894\sym{*}  &      0.0166\sym{**} &       0.862\sym{***}&     0.00106         &       0.360\sym{***}&     0.00850         &    -0.00150         \\
                    &     (59.25)         &      (1.12)         &     (40.32)         &      (2.15)         &      (3.20)         &     (86.05)         &      (0.29)         &     (43.54)         &      (1.70)         &     (-0.49)         \\
\midrule
Observations        &       14049         &       14049         &       14049         &       14049         &       14049         &       14049         &       14049         &       14049         &       14049         &       14049         \\
\midrule
Clustered SE & Yes & Yes & Yes & Yes & Yes & Yes & Yes & Yes & Yes & Yes \\ 
\bottomrule
\multicolumn{11}{l}{\footnotesize \textit{t} statistics in parentheses}\\
\multicolumn{11}{l}{\footnotesize \sym{*} \(p<0.05\), \sym{**} \(p<0.01\), \sym{***} \(p<0.001\)}\\
\end{tabular}
\end{adjustbox}
\end{table}


\section{Model estimates with White Standard Errors}\label{white_ses}

We report the DID model estimates with White standard errors robust to heteroskedasticy, assuming to within cluster correlation of the errors.

Once the estimated standard errors do not account for within-city correlations, we can no longer find a significant treatment effect on political uncertainty. The variance \emph{within} clusters is much different from the variance \emph{between} clusters, even as we correct for violation of homoskedastic error variance.

\begin{table}[H]\centering
\def\sym#1{\ifmmode^{#1}\else\(^{#1}\)\fi}
\caption{DiD Regression table for \emph{Uncertainty} and \emph{Negative Sentiment}, aggregated and grouped by topics with user fixed effects (omitted) with White Standard Errors.\label{user_2}}
\begin{adjustbox}{max width=\textwidth}
\begin{tabular}{lrrrrrrrrrr}
\toprule
&\multicolumn{5}{c}{Uncertainty}&\multicolumn{5}{c}{Negative Sentiment}\\
\toprule
                    &\multicolumn{1}{c}{(1)}&\multicolumn{1}{c}{(2)}&\multicolumn{1}{c}{(3)}&\multicolumn{1}{c}{(4)}&\multicolumn{1}{c}{(5)}&\multicolumn{1}{c}{(6)}&\multicolumn{1}{c}{(7)}&\multicolumn{1}{c}{(8)}&\multicolumn{1}{c}{(9)}&\multicolumn{1}{c}{(10)}\\
                    &\multicolumn{1}{c}{Aggregate}&\multicolumn{1}{c}{Economics}&\multicolumn{1}{c}{Health}&\multicolumn{1}{c}{Politics}&\multicolumn{1}{c}{Policy}&\multicolumn{1}{c}{Aggregate}&\multicolumn{1}{c}{Economics}&\multicolumn{1}{c}{Health}&\multicolumn{1}{c}{Politics}&\multicolumn{1}{c}{Policy}\\
\midrule
post=1              &      0.0378\sym{***}&     0.00819\sym{**} &      0.0592\sym{***}&    -0.00679\sym{*}  &      0.0198\sym{***}&     -0.0337\sym{***}&   -0.000346         &      0.0365\sym{***}&     -0.0108\sym{**} &     0.00736\sym{***}\\
                    &      (4.64)         &      (3.00)         &     (16.59)         &     (-2.50)         &      (9.40)         &     (-3.94)         &     (-0.12)         &     (11.74)         &     (-2.78)         &      (4.34)         \\
\addlinespace
post=2              &      0.0500\sym{***}&      0.0169\sym{***}&      0.0673\sym{***}&    -0.00253         &      0.0191\sym{***}&     -0.0405\sym{***}&     0.00614         &      0.0401\sym{***}&     -0.0159\sym{***}&     0.00783\sym{***}\\
                    &      (5.69)         &      (5.43)         &     (16.21)         &     (-0.87)         &      (8.25)         &     (-4.45)         &      (1.92)         &     (11.09)         &     (-3.97)         &      (4.50)         \\
\addlinespace
red zone=1 $\times$ post=1&       0.145\sym{***}&     0.00222         &      0.0606\sym{**} &      0.0158         &      0.0281\sym{*}  &     -0.0168         &     -0.0176         &      0.0614\sym{**} &      0.0281\sym{**} &      0.0389\sym{**} \\
                    &      (3.65)         &      (0.16)         &      (3.09)         &      (1.86)         &      (2.12)         &     (-0.40)         &     (-1.04)         &      (3.12)         &      (2.82)         &      (2.91)         \\
\addlinespace
red zone=1 $\times$ post=2&      0.0475         &    -0.00495         &     0.00294         &    -0.00647         &   -0.000174         &     -0.0224         &     -0.0115         &      0.0419\sym{*}  &      0.0358\sym{***}&      0.0233\sym{*}  \\
                    &      (1.11)         &     (-0.28)         &      (0.15)         &     (-0.81)         &     (-0.02)         &     (-0.49)         &     (-0.59)         &      (2.13)         &      (3.54)         &      (2.00)         \\
\addlinespace
Constant            &       0.902\sym{***}&     -0.0119         &     -0.0702\sym{***}&     0.00900         &     -0.0189         &      0.0629         &     0.00535         &     -0.0819\sym{***}&     -0.0199\sym{*}  &     -0.0311\sym{**} \\
                    &     (21.47)         &     (-0.69)         &     (-3.59)         &      (1.21)         &     (-1.67)         &      (1.42)         &      (0.28)         &     (-4.23)         &     (-2.14)         &     (-2.70)         \\
\midrule
Observations        &       28370         &       28370         &       28370         &       28370         &       28370         &       28370         &       28370         &       28370         &       28370         &       28370         \\
\bottomrule
\multicolumn{11}{l}{\footnotesize \textit{t} statistics in parentheses}\\
\multicolumn{11}{l}{\footnotesize \sym{*} \(p<0.05\), \sym{**} \(p<0.01\), \sym{***} \(p<0.001\)}\\
\end{tabular}
\end{adjustbox}
\end{table}

\end{document}